\documentstyle[12pt,epsf]{article}

\def\thefootnote{\fnsymbol{footnote}}

\newcommand{\eq}{\begin{equation}}
\newcommand{\en}{\end{equation}}
\newcommand{\eqa}{\begin{eqnarray}}
\newcommand{\ena}{\end{eqnarray}}

\newcommand{\var}{\varepsilon}

\newcommand{\br}{\langle}
\newcommand{\kt}{\rangle}
\newcommand{\lb}{\lbrack}
\newcommand{\rb}{\rbrack}

\newcommand{\th}{\theta}

\newcommand{\wh}{\widehat}

\newcommand{\lan}{\langle}
\newcommand{\ran}{\rangle}
\newcommand{\nonu}{\nonumber}

\newcommand{\txrm}{\textrm}
 
\newcommand{\dep}{\partial}

\def\mycaptionl#1{%
\refstepcounter{figure}
\begin{center}
\hskip 1pt\vskip -0.6cm
\begin{minipage}{14cm}
\small {\bf Fig. \hskip -3pt\arabic{figure}}: {\sl #1}
\end{minipage}
\null\hskip 1pt\vskip -0.2cm
\end{center}}

\newcommand{\JP}[1]{J.\ Phys.\ {\bf #1}}
\newcommand{\NP}[1]{Nucl.\ Phys.\ {\bf #1}}
\newcommand{\PL}[1]{Phys.\ Lett.\ {\bf #1}}

\newcommand{\PR}[1]{Phys.\ Rev.\ {\bf #1}}

\newcommand{\IJMP}[1]{Int.\ J.\ Mod.\ Phys.\ {\bf #1}}

\begin{document}
\begin{titlepage}
\begin{flushright}
DFTT 41/99\\
GEF-Th 5/99\\
\end{flushright}
\begin{center}
{\Large\bf Short distance behaviour of correlators  in the}
\vskip 0.3cm
{\Large\bf  2D Ising model in a magnetic field}
\end{center}
\vskip 0.8cm
\centerline{
M. Caselle$^a$\footnote{e--mail: caselle@to.infn.it},
P. Grinza$^a$\footnote{e--mail: grinza@to.infn.it} and
N. Magnoli$^b$\footnote{e--mail: magnoli@ge.infn.it}}
 \vskip 0.6cm
 \centerline{\sl  $^a$ Dipartimento di Fisica
 Teorica dell'Universit\`a di Torino and}
 \centerline{\sl Istituto Nazionale di Fisica Nucleare, Sezione di Torino}
 \centerline{\sl via P.Giuria 1, I-10125 Torino, Italy}
 \vskip .2 cm
 \centerline{\sl  $^b$ Dipartimento di Fisica,
 Universit\`a di Genova and}
 \centerline{\sl Istituto Nazionale di Fisica Nucleare, Sezione di Genova}
 \centerline{\sl via Dodecaneso 33, I-16146 Genova, Italy}
 \vskip 0.6cm

\begin{abstract}
We study the $\br\sigma\sigma\kt$, $\br\sigma\epsilon\kt$, 
$\br\epsilon\epsilon\kt$ correlators in the 2d Ising model perturbed by a
magnetic field. We compare the results of a set of high precision Montecarlo
simulations with the predictions of two different approximations: the Form
Factor approach, based on the exact S-matrix description of the model, and a
short distance perturbative expansion around the conformal point. Both methods
give very good results, the first one performs better for distances larger than
the correlation length, while the second one is more precise for distances
smaller than the correlation length. In order to improve this agreement 
we  extend the perturbative analysis to the second order in the 
derivatives of the OPE constants. 
 
\end{abstract}
\end{titlepage}

\setcounter{footnote}{0}
\def\thefootnote{\arabic{footnote}}

\section{Introduction}

In these last years much progress has been done in the study of two dimensional
statistical systems in the neighbourhood  of  critical points. In the framework
 of quantum field theory  these systems can be seen as Conformal Field Theories
 (CFTs) perturbed by some relevant operator. Since the seminal work of
 Belavin, Polyakov and Zamolodchikov~\cite{bpz} we have an almost complete
 understanding of 
CFT's (at least for the minimal models): we have  complete
 lists of all the operators of the theories and  explicit
 expressions for the correlators. However much less is known on their
 relevant perturbations. In some cases it has been possible to show that these
 perturbations give rise to integrable models~\cite{integrable,zam89}. 
 In these cases again we have a
 rather precise description of the theory. In particular it is possible to
 obtain the exact asymptotic expression for the large distance behaviour of the
 correlators~\cite{form}. {}From this information several important results 
 (and in particular all the universal amplitude ratios) can be obtained.

  However in comparing with numerical simulations or with experiments
  one is often
  interested in the short distance behaviour of the correlators (short here
  means for distances smaller or equal than the correlation length) and this is
  not easily accessible in integrable systems. Moreover integrable perturbations
  represent only a small subset of the possible theories. 
  For instance in the case of
  the Ising model both the purely thermal and the purely magnetic perturbations
  are integrable, but for any combination of them the exact
  integrability is lost.

For these reasons, besides the S-matrix results,
 it is important to develop a perturbative approach 
 well defined in the short distance
regime of the theory and such that it does not rely on the exact
integrability of the model.   This is however a rather difficult task. In fact
any naive perturbative expansion of  the
(massless) CFT  along a relevant direction, is affected by  
 infrared divergences (IR) and some non-trivial strategy is needed.

  Recently, in \cite{gm1,gm2}, a new approach has been proposed 
 to overcome this difficulty (see \cite{CPT}-\cite{david} for relevant 
related works and preexisting ideas).
 The method is based on Wilson's operator product expansion (OPE). 
 Roughly speaking the main idea of this new approach
 is that the Wilson's coefficients of the OPE, being well defined at short
 distance, 
 can be assumed to have a regular, IR safe, perturbative expansion with
 respect to the coupling.  For this
 reason we shall refer to it in the following as the IRS (infrared safe)
 perturbative approach.

 The main requirement of this IRS approach is the  the knowledge
 of the Wilson coefficients (and their derivatives with respect to the
 perturbing coupling). For this reason it is particularly efficient if applied
 to perturbations of exactly solved theories like 2d critical CFT's (but the
 framework is quite general and in principle could be extended also to
 higher dimensions).

  The price one has to pay to control in this way the IR divergences 
 is that one needs, as an external input information,
  the expectation values of the operators involved in the expansion.
  There are at this point two possibilities. The first one is to concentrate
  only on observables in which these expectation values exactly cancel. This
  is a small but very interesting 
  subset of the informations that we can obtain with the IRS
  perturbation.

  The second possibility is to obtain the desired expectation values 
   with some other method or extract them from numerical simulations (an
interesting numerical approach to obtain these VEV is based on the
Truncated Conformal Space technique, see \cite{yz,gm3} and references
therein).
 
  {}From this point of view, the IRS approach becomes
   particularly powerful if applied to integrable perturbations, since in this
   case some of the expectation values can be deduced from the S-matrix of the
   model. 

   The last step one has to face in comparing the results of the IRS method 
   with simulations or experiments is the presence of a nonuniversal
   normalization factor
   between the operators in the continuum quantum field theory and their lattice
   discretizations. These normalizations (and the related normalization of the
   coupling of the perturbation) can be fixed if an exact solution of the
   lattice model exists at the critical point. Actually much less is needed. One
   only needs the exact expression  (or even only its large distance 
   asymptotic form) of a correlator involving the operators in which we are
   interested. This makes the Ising model perturbed by a magnetic field a
   perfect candidate for testing the IRS method. In fact it is well known that
   this model is exactly integrable~\cite{integrable,zam89}
   and all the amplitude ratios and
   expectation values of the primary fields are known. Moreover the Ising model
   is exactly solvable at the critical point and the exact expression is known
   for several correlators~\cite{ising,wu}. 

In fact the IRS 
approach was successfully tested with the magnetic perturbation of
the Ising model in~\cite{gm2}. The aim of this paper is to make a
 further step in this
direction. In particular we have three goals:
\begin{description}

\item{a]}
Compare the results of the method with new high precision Montecarlo 
simulations so as to test the range of applicability of the
method.

\item{b]} Compare the IRS method with the results obtained in the 
S-matrix framework with the so called ``form factor'' (FF) approach.

\item{c]} 
Show that it is possible to extend the analysis of~\cite{gm1,gm2} 
to higher orders in the perturbative coupling 
and discuss the technical problems that one has to
afford following this route.

\end{description}

In particular we shall study in this paper, as an example, the second order 
term in the perturbative expansion of the $\br \epsilon \epsilon \kt$  
correlator. The reason for this choice is that this correlator  has a very 
peculiar behaviour since its first
order correction turns out to be exactly zero, thus
our second order calculation is mandatory  if one is interested to study
the influence of the magnetic field
 on the simple critical behaviour of the correlator.

This paper is organized as follows. Sect.~2 is devoted to a general
description of the Ising model in a magnetic field both on the lattice and in
the continuum. The aim of this section is
to fix conventions and normalizations which will be useful in the following.
In sect.~3 we shall briefly describe the IRS method, while in sect.~4 we shall
extend it to second order derivatives of the magnetic field. 
In sect.~5 we shall briefly
describe our Montecarlo simulation while in sect.~6 we shall compare the results
of our simulations with IRS and FF predictions. 
Finally sect.~7 will be devoted to some concluding remarks. The details of the
calculation of the second order derivative of the Wilson coefficient are
collected in the Appendix. We have reported in
three tables at the end of the paper a sample of the results of our simulations.

\section{Ising model in a magnetic field.}

The continuum theory, which is the starting point of the IRS expansion
is given by the action:
\eq
{\cal A} = {\cal A}_0 + h \int d^2x \, \sigma(x) \,\,\,
\label{action}
\en

where ${\cal A}_0$ is the action of the conformal field theory  
 which describes the Ising model at the critical point. Let us start our
 analysis by looking in detail at this CFT.

\subsection{The Ising model at the critical point}
The Ising model at the critical point is described by the 
unitary minimal CFT with central charge
$c=1/2$. It contains three conformal families whose
 primary fields ${\bf 1},\sigma,\epsilon$ have
scaling dimensions $x=0,1/8,1$ respectively. The fusion rule algebra is
\eq
\begin{array}{lll}
\lb \epsilon \rb \lb \epsilon \rb \ & = & \ \lb \mathbf{1} \rb \\
\lb \sigma\rb\lb\epsilon\rb \ & = & \ \lb\sigma\rb       \\
\lb\sigma\rb\lb\sigma\rb \ & = &\ \lb\mathbf{1}\rb \ + \ \lb\epsilon\rb.
\end{array}
\en

Once the operator content is 
known,
the only remaining information which is needed to completely
identify the theory are the OPE constants. The OPE algebra is defined as 
\eq
 \Phi_i(r) \Phi_j(0) =
          \sum_{ \{ k \}} C^{ \{ k \}}_{ij} (r) 
           \Phi_{\{ k \}}(0) 
\en
where with the notation $\{ k \}$ we mean that the sum runs over all the 
fields of the conformal family $[k]$. 
The structure functions $C^{k}_{ij} (r)$ are c-number functions of $r$ which
must be single valued in order to take into account locality. 
In the large $r$ limit they decay with  a power like behaviour
\eq
C^{k}_{ij} (r) \sim |r|^{-dim(C^{k}_{ij})}
\en
whose amplitude is given by
\eq
\hat C^{k}_{ij}~\equiv~\lim_{r\to\infty} C^{k}_{ij} (r)~|r|^{dim(C^{k}_{ij})} .
\label{u1}
\en

The actual value of these constants depends on the normalization of the fields,
which can be chosen by fixing the long distance behaviour of,
for instance, the $\sigma\sigma$ and $\epsilon\epsilon$ correlators. In this
paper we follow the commonly adopted convention which is\footnote{Notice 
the change of normalization
 with respect to~\cite{gm2}. The $\epsilon$ operator of the present paper 
corresponds to ${2\pi}$ times that of ref.~\cite{gm2}.}:
\eq
\br \sigma(x)\sigma(0) \kt =\frac{1}{\,\,|x|^{\frac{1}{4}}}\,\,,
\hspace{1cm}|x|\rightarrow
\infty
\label{uv}
\en
\eq
\br \epsilon(x)\epsilon(0)\kt=\frac{1}{\,\,|x|^{2}}\,\,,
\hspace{1cm}|x|\rightarrow
\infty.
\label{uve}
\en
With these conventions  
we have, for the structure constants among primary fields
\eq
\hat C^{\sigma}_{\sigma,\sigma}=\hat C^{\sigma}_{\epsilon,\epsilon}=
\hat C^{\epsilon}_{\epsilon,\sigma}=0
\label{f4}
\en
\eq
\hat C^{{\bf 1}}_{\sigma,\sigma}=\hat C^{\sigma}_{\sigma,{\bf 1}}=
\hat C^{{\bf 1}}_{\epsilon,\epsilon}=\hat C^{\epsilon}_{\epsilon,{\bf 1}}=1
\label{f4b}
\en
and
\eq
\hat C^{\sigma}_{\sigma,\epsilon}= 
\hat C^{\epsilon}_{\sigma,\sigma}=\frac12 .
\label{f4c}
\en

\subsection{The Ising model in a magnetic field}

If we switch on the magnetic field $h$, the structure functions acquire a $h$
dependence so that we have in general
\eq
 \Phi_i(r) \Phi_j(0)  =
          \sum_{ \{ k \}} C^{ \{ k \}}_{ij} (h,r) 
           \Phi_{ \{ k \}}(0) ~~~~.
\en

Also  the mean values of the $\sigma$ and $\epsilon$
operators acquire a dependence on $h$. 
Standard renormalization group arguments allow one to relate this $h$ dependence
to the scaling dimensions of the operators of the theory and lead to the
following expressions:
\eq
\langle\sigma\rangle_h=A_\sigma h^{\frac{1}{15}}+ ...
\label{mag}
\en
\eq
\langle\epsilon\rangle_h=A_\epsilon h^{\frac{8}{15}}+...
\en
The exact value of the two constants $A_\sigma$ and $A_\epsilon$
can be found in~\cite{fateev}  and \cite{flzz} respectively
\eq
A_\sigma=\frac{2\,{\cal C}^2}{15\,(\sin\frac{2\pi}{3}+
\sin\frac{2\pi}{5}+\sin\frac{\pi}{15})}=1.27758227..\,,
\label{sigmah}
\en
with
\eq
{\cal C} \,=\,
 \frac{4 \sin\frac{\pi}{5} \Gamma\left(\frac{1}{5}\right)}
{\Gamma\left(\frac{2}{3}\right) \Gamma\left(\frac{8}{15}\right)}
\left(\frac{4 \pi^2 \Gamma\left(\frac{3}{4}\right)
\Gamma^2\left(\frac{13}{16}\right)}{\Gamma\left(\frac{1}{4}\right)
\Gamma^2\left(\frac{3}{16}\right)}\right)^{\frac{4}{5}}~~  ,
\en
 and
\eq
A_\epsilon=2.00314...~~~.
\en
Notice however that these amplitudes are not universal. They depend on the
details of the  regularization scheme. Thus some further work is needed to
obtain their value on the lattice.

\subsection{The lattice model}
The lattice version of the above model is defined
by the following partition
function:

\eq
Z=\sum_{\sigma_i=\pm1}e^{\beta(\sum_{\br i,j \kt}\sigma_i\sigma_j+H\sum_i\sigma_i)}
\en
where the notation $\br i,j \kt$ 
denotes nearest neighbour sites in the lattice which we assume to be a two
dimensional square lattice of size $L$. 
In order to select only the magnetic perturbation, $\beta$ must be fixed to its
critical value:
 $$\beta=\beta_c=\frac12\log{(\sqrt{2}+1)}=0.4406868...$$

finally by defining  $h_l=\beta_c H$ we find
\eq
Z=\sum_{\sigma_i=\pm1}e^{\beta_c\sum_{\br i,j \kt }\sigma_i\sigma_j+h_l\sum_i\sigma_i}
\en
In the following we shall denote the
 lattice discretization of the operators $\sigma$, $\epsilon$ with the index
 $l$. 
The magnetization $M(h)$ is defined as usual:
\eq
M(h)\equiv\frac1N\frac{\partial}
{\partial h_l}(\log~Z)\vert_{\beta=\beta_c}
= \br \frac1N\sum_i \sigma_i \kt .
\en
where $N\equiv L^2$ denotes the number of sites of the lattice.
This result suggests  the following definition
for the lattice discretization of $\sigma$
\eq
\sigma_l\equiv \frac1N\sum_i \sigma_i ~~~,
\en
so that the mean value of $\sigma_l$ coincides with $M(h)$:
\eq
\br \sigma_l \kt \equiv M(h)
\en

Similarly, we define the internal energy as:
\eq
E(h) \equiv\frac{1}{2N}\frac{\partial}
{\partial h_l}(\log~Z)\vert_{\beta=\beta_c}
= \br \frac{1}{2N}\sum_{\br i,j \kt}\sigma_i\sigma_j \kt
\en
 For the energy operator one must also 
take into account the presence of an additional bulk contribution
at the critical point. This constant
 can be easily evaluated (for instance by
using Kramer-Wannier duality) to be $\epsilon_0=\frac{1}{\sqrt2}$. 
This result suggests, for the lattice discretization of $\epsilon$, 
 the following definition
\eq
 \epsilon_l  \equiv  \frac{1}{2N}\sum_{\br i,j \kt }\sigma_i\sigma_j
-\frac{1}{\sqrt{2}}
\label{bulk}
\en
so that the mean value of $\epsilon_l$ coincides with the singular part of
$E(h)$:
\eq
\br \epsilon_l \kt \equiv E(h)
-\frac{1}{\sqrt{2}}
\label{bulk2}
\en

According to the above discussion we expect:
\eq
\langle\sigma_l\rangle_h=A^l_\sigma h_l^{\frac{1}{15}}+ ...
\en
\eq
\langle\epsilon_l\rangle_h=A^l_\epsilon h_l^{\frac{8}{15}}+...
\en
where the lattice amplitudes $A^l_\sigma$, $A^l_\epsilon$ are
  different from the corresponding amplitudes evaluated in the
continuum.

In order to relate the lattice results with the continuum ones, we must fix the
relative normalizations of $\sigma$ versus $\sigma_l$ , 
 $\epsilon$ versus $\epsilon_l$ and $h$ versus $h_l$\footnote{This essentially
 amounts to measure all the quantities in units of the lattice spacing. For this
 reason we can fix in the following the lattice spacing to 1 and neglect it.}.
 The simplest way to do
this is to look at the analogous of eq.(\ref{uv},\ref{uve}) at 
 the critical point (namely for $h_l=0$)~\cite{smilga}. 
{}From the exact solution of the Ising model~\cite{ising} we know that
\eq
 \label{def2}
\br \sigma_i \sigma_j \kt_{h=0}\ = \ \frac{ R_{\sigma}^2}
{|r_{ij}|^{1/4}}
\en
where  $r_{ij}$ denotes the distance on the lattice between the sites $i$ and
$j$ and we know from \cite{wu} that:
\eq
R^2_{\sigma}=e^{3\xi'(-1)}2^{5/24}=0.70338...
\en

By comparing this result with eq.(\ref{uv}) we find
\eq
 \sigma_{l} \ = \ R_{\sigma}  \sigma= 0.83868...
   \sigma
\label{rss}
\en

{}From this we can also obtain the normalization of the lattice 
magnetic field which must exactly compensate that of the spin operator in the 
perturbation term 
$h\sigma$.
We find:

\eq
 h_{l} \ = \ (R_{\sigma})^{-1}  h= 1.1923... h
\label{ssh}
\en

Combining these two results we obtain the value in lattice units of the constant
$A_\sigma$
\eq
A_\sigma^l=(R_\sigma)^{16/15}A_\sigma=1.058...
\label{magl}
\en
In the case of the energy operator the connected correlator on the lattice, 
at $h_l=0$ and for any value of $\beta$, has the following
expression~\cite{h67}:
\eq
\br \epsilon_l(0)\epsilon_l(r) \kt_c=\left(\frac{\delta}{\pi}\right)^2
\left[ K_1^2(\delta r)- K_0^2(\delta r)
\right]
\en
where $K_0$ and $K_1$ are modified Bessel functions, $\delta$ is a parameter
related to the reduced temperature, defined as
\eq
\delta=4|\beta-\beta_c|
\en
and with the index $c$ we denote the connected correlator (notice that thanks to
the definition (\ref{bulk}) no disconnected part must be  subtracted at the
critical point and the index $c$ becomes redundant). This expression has a
finite value in the $\delta\to0$ limit (namely at the critical point). In fact
the Bessel functions difference can be expanded in the small argument limit as
\eq
\left[ K_1^2(\delta r)- K_0^2(\delta r)
\right]=\frac{1}{(\delta r)^2}+...
\en
thus giving, exactly at the critical point:
\eq
\br \epsilon_l(0)\epsilon_l(r)\kt =\frac{1}{(\pi r)^2}\equiv 
\frac{ R_{\epsilon}^2}
{|r|^{2}} .
\label{defe}
\en
Hence $R_\epsilon=1/\pi$. By comparing this result with eq.(\ref{uve}) we find
\eq
 \epsilon_{l} \ = \ R_{\epsilon}  \epsilon=  \frac{\epsilon}{\pi}
\label{ss2}
\en
and from this we obtain the expression in lattice units of $A_\epsilon$
\eq
A_\epsilon^l=(R_\sigma)^{8/15}(R_\epsilon)A_\epsilon=0.58051...
\en

\subsection{Correlators}
In the remaining part of this paper we shall be mainly interested in the  
dependence on the external magnetic field of the
following correlators:
\eq
G_{\sigma,\sigma}\equiv \br \sigma(0)\sigma(r) \kt 
\en
\eq
G_{\epsilon,\epsilon}\equiv \br \epsilon(0)\epsilon(r) \kt
\en
\eq
G_{\sigma,\epsilon}\equiv k \br \sigma(0)\epsilon(r)\kt
\en
where $k\equiv sign(h)$.
We already know the behaviour at the critical point of the first two of them, 
which is given by eq.
(\ref{uv}), (\ref{uve}) in the continuum (or equivalently 
eq.(\ref{def2}), (\ref{defe}) on the lattice), 
 while the OPE constants reported in eq.(\ref{f4}) immediately 
tell us that $\langle \epsilon(0)\sigma(r) \rangle =0$.

For small values of $h$ we may expect to add to these results 
 correction terms functions of $h$ and $r$.
However standard 
renormalization group arguments show that these two variables are actually
related and that there is
 a natural scaling variable
which describes the short distance expansion
 of these correlators in a magnetic field which is $t\equiv|h|~|r|^{15/8}$.
In order to obtain an explicit expansion in powers of $t$ we must absorb the
scaling dimensions of the various operators in the expansion~\footnote{
On the contrary, in the large distance regime where one may use the predictions
obtained with the form factor approach, the natural normalization is that
of  the $G_{\Phi,\Phi}$ functions defined above.}. 
 To this end let us
define
\eq
F_{\sigma,\sigma}\equiv \br \sigma(0)\sigma(r) \kt |r|^{1/4}
\en
\eq
F_{\epsilon,\epsilon}\equiv \br \epsilon(0)\epsilon(r)\kt |r|^{2}
\en
\eq
F_{\sigma,\epsilon}\equiv k \br \sigma(0)\epsilon(r) \kt |r|^{9/8}
\en
where $k\equiv sign(h)$.

 The powers which appear in the $t$ expansion of the functions $F$ 
can be immediately deduced from the analysis of the OPE via the IRS
method, that will be described in the following section.

\section{The infrared safe approach}

The goal of the method presented in Ref.  \cite{gm1} is to obtain informations
about the short distance behavior of a conformal field theory perturbed by
relevant operators.

The general idea behind this approach,
(for preexisting ideas see ~\cite{zamo,sonoda1,mz,wilson,wegner})
is the fact that Wilson Coefficients, being short distance 
objects, can be taken to have a regular, IR safe, 
perturbative expansion with respect to the coupling.
This OPE approach leaves unfixed some constants that parameterize the 
vacuum expectation values
of operators that appear in conformal field theory.

In \cite{gm1} it was found that the correlators of the perturbed CFT are
given in terms of the derivatives of the Wilson coefficients (calculated at
$h=0$ point). To be precise, they appear in the following way
\eqa
\lefteqn{ \lan \Phi_i(r) \Phi_j(0) \ran_h =
          \sum_{ \{ k \}} C^{ \{ k \}}_{ij} (h,r) 
          \lan \Phi_{ \{ k \}}(0) \ran_h= {}}
                                                \nonu\\
 & & {}   \sum_{ \{ k \}} \left\lb C^{ \{ k \}}_{ij} (0,r) +
          \dep_h C^{ \{ k \}}_{ij} (0,r) \  h  +
          \frac{1}{2} \dep_h^2 C^{ \{ k \}}_{ij} (0,r) \  h^2 +
          \cdots \right\rb
          \lan \Phi_{ \{ k \}}(0) \ran_h. 
\nonu
\ena
It was also shown that a general formula could be written for the $n-th$ 
derivative of the Wilson Coefficients with respect to $h$.
Here we will write only the first and the second order derivatives for the Wilson Coefficients, 

\eqa
\sum_b \dep_h^1 C^b_{a_1 a_2} \lan \phi_b X_R \ran  & = &
\nonu\\
 & = & \int' \txrm{d}^2 z   
 \lan \lb  \sigma_z  \left( \phi_a \phi_b - \sum_b^* 
 C^b_{a_1 a_2} \phi_b
 \right) X_R
 \rb \ran
\label{first}
\ena
and
\eqa
\sum_b \dep_h^2 C^b_{a_1 a_2} \lan \phi_b X_R \ran  & = &
\nonu\\
 & = & \int' \txrm{d}^2 z   \int' \txrm{d}^2 z' 
 \lan \lb  \sigma_z \sigma_{z'} \left( \phi_a \phi_b - \sum_b^* 
 C^b_{a_1 a_2} \phi_b
 \right) X_R
 \rb \ran  +
\nonu\\
 & + & \sum_b^* \dep_h C^b_{a_1 a_2} \int' \txrm{d}^2 z
 \lan \lb \sigma_z \phi_b X_R \rb \ran. 
\label{second}
\ena 
The general structure of this formula is a sum of the ``naive'' perturbative term plus $n$ (for the $n-th$ order coefficient)
infrared counterterms. The asterisk reminds that the 
sum on the counterterms is truncated and is performed up to a given infrared 
dimension (see again ~\cite{gm1}). \\

The construction of the IRS expansion requires two steps. 
\begin{itemize}
\item
First, one must
select by using the OPE rules which operators can appear
 in the various expansions, identify their scaling dimensions, select the
 dominant ones and give their expression in terms of the structure constants
 and of their derivatives.  
\item 
Second, one must evaluate the derivatives of the structure constants by
using eq.(\ref{first}) or (\ref{second}) to reduce them to suitable 
integrals over correlators  evaluated {\sl at the critical point}. This 
allows in principle to complete the analysis, since 
the explicit form of all possible critical correlators is known. However in
general these integrals are highly non trivial and their evaluation 
represents the major problem of the whole approach.

\end{itemize}

These two steps where performed in~\cite{gm2} for all the terms in the expansion
involving at most first order derivatives of the structure constants. This
allows to obtain the first three terms in the expansion of the 
$\lan \sigma \epsilon \ran$ and $\lan \sigma \sigma \ran$ correlators (which
are reported for completeness at the end of this section).
On the contrary for the $\lan \epsilon \epsilon \ran$ in this way one can only 
obtain the first two terms of the expansion. 
Moreover one can verify by an explicit calculation that the second of them is
identically zero (\cite{gm1}). 
Thus in the $\lan \epsilon \epsilon \ran$ correlator, in order to reach the
first non trivial correction to scaling, it is mandatory to extend the analysis
of \cite{gm2} and to 
 deal with second order derivatives of the Wilson coefficients.  
We shall address this problem in the next section. In particular, in sect.~4.1
we shall discuss the first step of the IRS analysis, and select among the
possible candidates the one with the lowest power of $t$ which, as anticipated,
turns out to involve a second order derivative of a structure constant. Then
in sect.~4.2 (and in the Appendix) we shall explicitly evaluate this
contribution. Let us conclude this section by listing for all the three
correlators the first three terms of the IRS expansion 

\eq
F_{\sigma, \sigma}= B_{\sigma\sigma}^1 + B_{\sigma\sigma}^2 t^{8/15}
+ B_{\sigma\sigma}^3 t^{16/15}+O(t^2) \label{ss}
\en
\eq
F_{\epsilon ,\epsilon}= B_{\epsilon\epsilon}^1
+B_{\epsilon\epsilon}^2 t^{16/15}+
B_{\epsilon\epsilon}^3 t^{2}
+O(t^{32/15})\label{ee}
\en
\eq
\kappa
F_{\sigma , \epsilon}  = B_{\sigma\epsilon}^1 t^{1/15}
             +B_{\sigma\epsilon}^2 t+
B_{\sigma\epsilon}^3 t^{23/15}+
  O(t^{31/15}) \label{se}
\en
where the coefficients $B^i_{\Phi \Phi}$ are
\eqa
B^{1}_{\sigma \sigma} & = & \wh{C^{\mathbf 1}_{\sigma \sigma}} \nonu \\
B^{2}_{\sigma \sigma} & = & A_\epsilon \wh{C^{\epsilon}_{\sigma \sigma}}\nonu \\
B^{3}_{\sigma \sigma} & = & A_\sigma \wh{\dep_h C^\sigma_{\sigma \sigma}} \nonu\\
B^{1}_{\epsilon \epsilon} & = & \wh{C^{\mathbf 1}_{\epsilon \epsilon}} \nonu\\
B^{2}_{\epsilon \epsilon} & = & A_\sigma \wh{\dep_h C^\sigma_{\epsilon
\epsilon}} \nonu\\ 
B^{3}_{\epsilon \epsilon} & = & \frac{1}{2}  \wh{\dep_h^2 C^{\mathbf 1}_
{\epsilon \epsilon}}\nonu \\
B^{1}_{\sigma \epsilon} & = & A_\sigma \wh{C^\sigma_{\sigma \epsilon}} \nonu\\
B^{2}_{\sigma \epsilon} & = & \wh{\dep_h C^{\mathbf 1}_{\sigma \epsilon}} \nonu\\
B^{3}_{\sigma \epsilon} & = & A_\epsilon \wh{\dep_h C^\epsilon_{\sigma\epsilon}}
\nonu\ena
 the derivatives of $C_{\Phi_i \Phi_j}^{\Phi_k}$ are given by (\ref{first}) and
 the notation $\wh{\dep_h C_{\Phi_i \Phi_j}^{\Phi_k}}$ is the extension of the
 definition given in eq.(\ref{u1}) to the derivatives of the Wilson
 coefficients. The first order derivatives have been calculated in~\cite{gm2}
 we report here for completeness their numerical value. Notice a slight change
 with respect to~\cite{gm2} due to the different choice of normalizations for
 the $\epsilon$ operator (the $\epsilon$ of the present paper corresponds to
 ${2\pi}$ times that of ref.~\cite{gm2})  

\eqa
 \wh{\dep_h C^\sigma_{\sigma \sigma}}&=& -0.40374\nonu\\
 \wh{\dep_h C^\sigma_{\epsilon\epsilon}}&=& 0\nonu\\ 
 \wh{\dep_h C^{\mathbf 1}_{\sigma \epsilon}}&=&3.29627 \nonu\\
 \wh{\dep_h C^\epsilon_{\sigma\epsilon}} &=& -0.90900
\nonu\ena

 The second order derivative which appears in
last one in the $\lan \epsilon \epsilon \ran$ correlator requires a more
involved calculation which we shall  discuss  in the next section and in the
Appendix.

\section{Second order corrections}
\subsection{Dimensional analysis}
To estimate  the higher order corrections to  $\lan \epsilon \epsilon \ran$ we must analyse two
kinds of possible contributions.
\begin{itemize}
\item
The expectation values of secondary operators multiplied by the Wilson coefficients and their first 
derivatives.
\item
 The second derivatives of Wilson coefficients.
\end{itemize}
We would like to understand which is the most important of these terms.
In the Ising model there are two secondary operators at first level, obtained by acting on $\epsilon$ and $\sigma$ with 
the Virasoro generator $L_{-1}$  and its 
hermitian conjugate (the action of $L_{-1}$ on $\mathbf{1}$ 
gives $0$).
We  start by considering
\eq
\epsilon^1 \ \equiv \ L_{-1} {\bar L}_{-1} \epsilon 
\en
and 
\eq
\sigma^1 \ \equiv \ L_{-1} {\bar L}_{-1} \sigma
\en
where  $L_k$, $\bar L_k$ are Virasoro generators.
It is clear that the expectation value of this kind of operators is zero being total derivatives.
So let us go to second level of the algebra. There are two possible terms:
\eq
L_{-1}^2 {\bar L}^{2}_{-1} \phi, \ \ \ \ \ \ \ \ \ \ \ \ \ \ \ 
L_{-2} {\bar L}_{-2}  \phi
\en
where $\phi$ is a generic primary field. In this situation we have to consider also the identity operator.

In the identity sector the only contribution is given by
\eq
T \bar T \ = \ L_{-2} {\bar L}_{-2} \ \mathbf 1
\en
i.e. the energy-momentum tensor. Again a simple dimensional analysis shows that
\eq
\txrm{dim} \ T \bar T \ = 4, \ \ \ \ \ \ \ \ \ \ \ 
\lan T \bar T \ran_h \ = \ A_{T \bar T} \vert h \vert^{32/15}
\en
giving 
\eq
A_{T \bar T} \ \wh{C^{T \bar T}_{\epsilon \epsilon}} \ t^{32/15}.
\en
It is clear that the terms containing secondary operators (of second level) of $\sigma$ and $\epsilon$
are of higher order in $t$ and will not be considered here.

A second possible contribution is given by the higher order derivative
\eqa
\sum_b \dep_h^2 C^b_{a_1 a_2} \lan \phi_b X_R \ran  & = &
\nonu\\
 & = & \int' \txrm{d}^2 z   \int' \txrm{d}^2 z' 
 \lan \lb  \sigma_z \sigma_{z'} \left( \phi_a \phi_b - \sum_b^* 
 C^b_{a_1 a_2} \phi_b
 \right) X_R
 \rb \ran  +
\nonu\\
 & + & \sum_b^* \dep_h C^b_{a_1 a_2} \int' \txrm{d}^2 z
 \lan \lb \sigma_z \phi_b X_R \rb \ran. 
\ena
Let us fix $X_R =\mathbf 1$.  An elementary computation shows
that the series is truncated and only those operators having $x_b \leq 
\frac{15}{8}$ appear in it.

It follows that
\eqa
\dep_h^2 C^{\mathbf{1}}_{\epsilon \epsilon} & = &
\nonu\\
 & = & \int' \txrm{d}^2 z  \int' \txrm{d}^2 z' 
 \lan \sigma_z \sigma_{z'} (\epsilon_r \epsilon_0 - 
 C^{\mathbf{1}}_{\epsilon \epsilon} ) \ran    +
\nonu\\
 & + & \dep_h C^\sigma_{\epsilon \epsilon} 
   \int' \txrm{d}^2 z  \lan \sigma_z  \sigma_0 \ran
\ena
but  $\dep_h C^\sigma_{\epsilon \epsilon}=0$, and we can say that
\eq
\dep_h^2 C^{\mathbf{1}}_{\epsilon \epsilon} 
  =  \int' \txrm{d}^2 z  \int' \txrm{d}^2 z' 
 \lan \sigma_z \sigma_{z'} (\epsilon_r \epsilon_0 - 
 C^{\mathbf{1}}_{\epsilon \epsilon} ) \ran.    
\en

Again from dimensional analysis we get
\eq
\txrm{dim} \ \dep_h^2 \wh{C^{\mathbf{1}}}_{\epsilon \epsilon} \ = \  
\frac{14}{8}
\en
and the contribution to $\lan \epsilon \epsilon \ran$ of the second
order derivative is given by
\eq
\frac{1}{2} \ \dep_h^2 C^{\mathbf{1}}_{\epsilon \epsilon}  \ t^2.
\en

It is also clear that, being $X_R=\mathbf 1$ the lowest dimension
operator,  derivatives of Wilson coefficients relative to $\sigma$
and $\epsilon$ will give terms with a higher power in $t$.

Let us write finally the perturbative expansion of this  correlator
\eq
F_{\epsilon \epsilon}  \ = \ 
\wh{C^{\mathbf{1}}_{\epsilon \epsilon}} \ + \ 
\frac{1}{2} \ \dep_h^2 \wh{C^{\mathbf{1}}_{\epsilon \epsilon}}  \ t^2 + 
O(t^{32/15}).
\en

\subsection{The Wilson derivative}
Let us remember that 
\eq
\dep_h^2 C^{\mathbf{1}}_{\epsilon \epsilon} 
  =  \int' \txrm{d}^2 z  \int' \txrm{d}^2 z' 
 \lan \sigma_z \sigma_{z'} (\epsilon_r \epsilon_0 - 
 C^{\mathbf{1}}_{\epsilon \epsilon} ) \ran    
\en
where $\lan \sigma(z_1) \sigma(z_2) \epsilon(z_3) \epsilon(z_4) \ran$ denotes
the correlator at the critical point which
 can be written as
\eq
\lan \sigma(z_1) \sigma(z_2) \epsilon(z_3) \epsilon(z_4) \ran =
\frac{\vert z_{12}(z_{32}+z_{42})-2z_{32}z_{42}\vert^2}
{4  \vert z_{42}z_{32}z_{41}z_{31} \vert \vert z_{43} \vert^2 
\vert z_{12} \vert^{1/4}}.
\en

By fixing the values of 
$z_1=z$, $z_2=w$, $z_4=0$ and by  rescaling  $r$ we can 
choose  $z_3=1$, we get
\eq
\dep_h^2 C^{\mathbf{1}}_{\epsilon \epsilon} 
  =  \int' \txrm{d}^2 w  \int' \txrm{d}^2 z 
\frac{\vert z(1-w)+w(1-z) \vert^2}
{4 \vert w(1-w) z(1-z) \vert 
\vert z-w \vert^{1/4}} + \cdots
\en
where the dots indicate the counterterms.

 The explicit calculation
of this integral can be done using a technique developed by
Mathur, \cite{mathu}. The general idea behind this approach is 
to factorize the integral in a holomorphic and antiholomorphic
part using Stokes theorem. The calculation is reported in the Appendix.
After this calculation, in order to get rid of the 
infrared cutoff we  perform a Mellin transform of the integral
and the infrared counterterm (see \cite{gm2} for more details). 
In this we we end up with a finite result when the infrared
cutoff goes to infinity. The final result is
\eq
\wh{\dep_h^2 C^{\mathbf{1}}_{\epsilon \epsilon}} \ = \
97.5936 \dots.
\en

Let us stress that the techniques that we have discussed in this section
 can be extended  to any order in the derivatives of
the Wilson coefficients. This is a rather important observation,
 since it allows, in principle, to study the IRS
corrections, in a consistent way, to any given order in $t$.

\section{The Montecarlo simulation}
It has been recently shown~\cite{ch} that in the case of  
 the 2d Ising model in a magnetic field, 
algorithms based on the exact (or approximate)
diagonalization of the transfer matrix are much more effective than standard
Montecarlo simulations. In particular this is true for  all
possible observables involved in the large distance behaviour of the model.
The only exception is represented by
the short distance behaviour of the point-point correlators 
which is  the subject of the present paper. In fact in order 
to reach lattices as large as possible in the 
 transfer-matrix programs discussed in~\cite{ch}
only zero momentum projected observables could be studied,
while we are instead interested in point-point correlators. 
 Moreover, we need to have a window as large as possible between the region (few
 lattice spacings) dominated  by the lattice artifacts and the correlation
 length. This windows shrinks to zero in the transfer matrix approach where 
 only small values of the correlation length can be studied.

 For these reasons we decided to perform our tests with  standard Montecarlo
 simulations.  We used a  Swendsen-Wang type algorithm,
modified so as to take into account the presence of an external magnetic field.
For a detailed description of the algorithm see for instance~\cite{lr,ddort}.

\subsection{Finite size effects.}
As a preliminary test we performed a simulation at $h_l=4.4069\times10^{-4}$ 
(which corresponds
to $H=0.001$) with lattice size $L=128$ which exactly coincides with one of 
 the simulations reported in~\cite{lr} and found results in complete agreement
 with those quoted in~\cite{lr}.
Then for the same value of $h_l$
we performed a set of high precision simulations varying the lattice size
so as to check the presence of possible finite size effects.
 In particular we compared
our estimates of the mean magnetization, susceptibility and internal energy 
  with the known exact
results, extrapolated at the value of $h_l$ at which we performed the 
simulations~\footnote{If one is interested in a high precision comparison, also
the contribution of secondary fields should be taken into account in extracting
these exact estimates. The amplitude of some of these secondary fields have been
evaluated numerically in~\cite{ch}. In the case of $M$ and $\chi$,
 for the values of $h_l$ in
which we are interested, the contributions of the secondary 
fields are strictly
smaller than the statistical errors of the results of our simulations and hence
can be neglected in the comparison. On the contrary for the internal energy it
turns out that the amplitude of the first correction 
is rather
large (see~\cite{ch} for details)
 and must be taken into account. In fact, if we would
neglect it, instead of the value reported in tab.1 we would find 
$E=0.71652$ in clear disagreement with the Montecarlo results. This represents a
non trivial test of the results of~\cite{ch}.}.

\begin{table}[h]
\label{tab1}
\caption{\sl Finite size effects at $h_l=4.4069\times 10^{-4}$. 
In the first column we report
the lattice sizes used in the simulations, in the remaining three columns the 
mean values of the magnetization, susceptibility and internal energy.
In the last row we report the exact results obtained by using the known
values of the amplitudes for these quantities.}
\vskip 0.2cm
\begin{tabular}{|c|l|l|l|}
\hline
$L$ &$M$  & $\chi$ & $E$     \\
\hline
120 & 0.63110(16) & 113(1)  & 0.71638(3)  \\
140 & 0.63247(16) & 98.4(6) & 0.71645(3)   \\
160 & 0.63245(14) & 96.4(4) & 0.71639(3)   \\
200 & 0.63255(11) & 95.4(3) & 0.71643(3)   \\
exact & 0.63260 & 95.7 & 0.71642   \\
\hline
\end{tabular}
\end{table}
The comparison is reported in tab.1. It turns out that lattice sizes
at least larger than 12 times the correlation length are needed to 
be sure that finite size
effects are  under control (with this we mean that the systematic errors 
induced by the finite size of the lattice are smaller than the statistical 
errors of the simulations and can be neglected). 
A side consequence of this
observation is that the simulations reported in~\cite{lr} are indeed affected 
by rather large finite size effects.

It is interesting to notice
that the magnetic observables are more affected by finite size effects than the
thermal one. As one can easily expect the largest corrections appear in the
case of the susceptibility. 

\subsection{The simulation}
Once we were sure to have finite size effects under control we performed a set
of high precision simulations
 of the model for three
different values of the magnetic field.

An important quantity to understand the range of validity of the IRS
approximation is the correlation length. Roughly speaking we expect that the IRS
results should give a reasonable approximation for distances equal or smaller
than
the correlation length, while above it the form factor approach should give
results of better quality. For this reason it is important to have a good
estimate of $\xi$. 
This can be easily obtained from the knowledge of the spectrum of the theory.
We find, in lattice units:
          
\eq
\label{xi}
\xi(h_l)=0.24935... h_l^{-\frac{8}{15}}
\en

 (see~\cite{ch} for details on the 
 continuum to lattice conversion of $\xi$).
In tab. 2 we have reported the expected values of $\xi$ in our cases.

\begin{table}[h]
\label{tab2}
\caption{\sl Values of the correlations lengths for the three choices of $h_l$.}
\vskip 0.2cm
\begin{tabular}{|c|c|}
\hline
$h_l$ &$\xi$    \\
\hline
$4.4069\times10^{-4}$ & 15.4 \\
$2.2034\times10^{-4}$ & 22.4 \\
$1.1017\times10^{-4}$ & 32.2 \\
\hline
\end{tabular}
\end{table}

For all the values of $h_l$ that we simulated, we studied the three correlators:
$\langle \sigma(0)\sigma(r) \rangle$,
$\lan \sigma(0)\epsilon(r) \ran$ and
$\lan \epsilon(0)\epsilon(r) \ran$,
for $r=1,\cdots,L_{max}$, where the maximum distance $L_{max}$ was chosen
 to be roughly twice the correlation length. In this way we can test our
 results also in the large distance regime, where predictions from
 the form  factor approach are expected to give very precise estimates for the
 correlators. Notice that when studying the large distance 
 behaviour of
 correlators one is usually interested in the zero momentum projection of the
 $connected$ part of the correlator. On the contrary in the present case we are
 interested  
 in the {\sl point--point} correlators without  mean value
 subtraction or zero momentum projections. This must be taken into account when
 comparing the data with those obtained with the form factor approach.
Some informations on the simulations are reported in tab.3.

\begin{table}[h]
\label{tab3}
\caption{\sl Some informations on the simulations. $L_{max}$ denotes the maximum
distance at which the correlators have been evaluated, its value almost
coincides with twice the correlation length. $L$ denotes the lattice size, $h_l$
the magnetic field. In the third column we have reported the number of measures
while in the fourth column we have reported the number of SW
sweeps which separates two measures. }
\vskip 0.2cm
\begin{tabular}{|c|c|c|c|c|}
\hline
$h_l$ &$L$  & measures & sweep/measures  & $L_{max}$    \\
\hline
$4.4069\times10^{-4}$ & 200 & $4\times 10^5$& 5  & 30 \\
$2.2034\times10^{-4}$ & 300 & $2\times 10^5$& 5  & 45 \\
$1.1017\times10^{-4}$ & 400 & $1\times 10^5$& 10  & 65 \\
\hline
\end{tabular}
\end{table}

We report an example of our results (for the value $h_l=4.4069\times10^{-4}$)
in the first
columns of tabs. 7, 8 and 9. The quoted errors have been obtained with 
a standard jacknife method.

\section{Discussion of the results}
In fig.s 1-8 and tab.7, 8 and 9 we compare our estimates for the correlators
with the IRS and form factor predictions. For completeness we briefly recall
here the form factor results (see~\cite{dm,ds} for details) and give the
numerical values (once all the conversion factors are taken into account) of the
constants in the IRS approach.

\subsection{Form factor results}

The scattering theory which describes the scaling limit of the Ising Model in a
magnetic field~\cite{zam89}
 contains eight different species of self-conjugated particles
$A_{a}$, $a=1,\ldots,8$ with masses
\eqa
m_2 &=& 2 m_1 \cos\frac{\pi}{5} = (1.6180339887..) \,m_1\,,\nonumber\\
m_3 &=& 2 m_1 \cos\frac{\pi}{30} = (1.9890437907..) \,m_1\,,\nonumber\\
m_4 &=& 2 m_2 \cos\frac{7\pi}{30} = (2.4048671724..) \,m_1\,,\nonumber \\
m_5 &=& 2 m_2 \cos\frac{2\pi}{15} = (2.9562952015..) \,m_1\,,\\
m_6 &=& 2 m_2 \cos\frac{\pi}{30} = (3.2183404585..) \,m_1\,,\nonumber\\
m_7 &=& 4 m_2 \cos\frac{\pi}{5}\cos\frac{7\pi}{30} = (3.8911568233..) \,m_1\,,
\nonumber\\
m_8 &=& 4 m_2 \cos\frac{\pi}{5}\cos\frac{2\pi}{15} = (4.7833861168..) \,m_1\,
\nonumber
\ena
$m_1(h_l)$ denotes
the overall mass scale and coincides with $1/\xi$, hence
its value in lattice units is
\eq
m_1(h_l) \,=\, 4.0104... \, h_l^{\frac{8}{15}} .
\en

From the knowledge of the masses and of the form factors
 it is possible to obtain a large distance
approximation for the correlators by constructing
a spectral sum over a complete set of
intermediate states.  We thus find for any pair of local
 operators $\Phi_1$ and $\Phi_2$:

\begin{eqnarray}
G_{\Phi_1\Phi_2}(x)&\equiv&\langle\Phi_1(x) \Phi_2(0)\rangle\nonumber\\
& = & \sum_{n=0}^{\infty}\int_{\th_1 >\th_2 \ldots>\th_n} 
\frac{d\th_1}{2\pi} \cdots \frac{d\th_n}{2\pi}\,\,
F^{\Phi_2*}_{a_1\ldots a_n}(\th_1,\ldots,\th_n)
F^{\Phi_1}_{a_1\ldots a_n}(\th_1,\ldots,\th_n)\nonumber\\
& \times & e^{-|x| \sum_{k=1}^n m_k \cosh\th_k}\,\,,
\label{ag1} 
\end{eqnarray}
where the form factors $F^{\Phi}_{a_1\ldots a_n}(\th_1,\ldots,\th_n)$ are the
matrix elements of the local operator $\Phi(x)$ on the asymptotic states
$A_{a}$, i.e they are defined as
\eq
F^{\Phi}_{a_1\ldots a_n}(\th_1,\ldots,\th_n) = \langle 0|
\Phi(0)|A_{a_1}(\th_1)\ldots A_{a_n}(\th_n)\rangle\,\,\,.
\label{ag2}
\en

The important point is that these form factors can be exactly 
 computed in the integrable models once the $S$--matrix is
known.

It is  natural to organize the above
 expansion  setting a reference value $m_r$ and keeping in the spectral sum only
 the states with a mass smaller than $m_r$. Looking at eq.(\ref{ag1}) we see
 that we must expect, as a consequence of this truncation, a
 systematic error in the approximation of the order 
${\cal O}\left(e^{-m_rx}\right)$.
 Up to $m_r=2m_1$ (where $m_1$ is the
 fundamental mass of the model) only single particle states survive in the sum
 and eq.(\ref{ag1}) greatly simplifies. In particular looking at eq.(66) we see
 that only the first three
 states (which are the only ones below the pair production threshold)
 survive. 
 Eq.(\ref{ag1}) becomes in this case, (choosing for instance
 the spin-spin correlator)
\eq
{G_{\sigma\sigma}(r)}\sim M^2(h_l)(1+\sum_{i=1}^3
\frac{\left|F^\sigma_i\right|^2}{\pi} K_0(m_i r))
\label{ffss}
\en
where $K_0(x)$ is the modified Bessel function and
$F^\sigma_i$ denotes the overlap (measured in units of the magnetization)
of the $i^{th}$ state with the $\sigma$ operator.
 $M(h_l)$ denotes the magnetization and its $h_l$ dependence is given
in (\ref{mag}, \ref{magl}).

 Similarly the spin-energy and energy-energy correlators are given by
\eq
k{G_{\sigma\epsilon}(r)}\sim M(h_l)E(h_l)(1+\sum_{i=1}^3
\frac{F^{\sigma *}_iF^\var_i}{\pi} K_0(m_i r))
\label{ffes}
\en
\eq
{G_{\epsilon\epsilon}(r)}\sim E^2(h_l)(1+\sum_{i=1}^3
\frac{\left|F^\var_i\right|^2}{\pi} K_0(m_i r))
\label{ffee}
\en

The constants $F^\sigma_i$ and $F^\var_i$ have been evaluated 
in~\cite{dm,ds}. Their value is reported in tab.4 and 5.

One can systematically improve the
approximation by setting higher values of $m_r$. For instance, for $m_r=3$ one
must keep into account the first five single particle form factors
 together with
 the two-particle ones involving the (1,1), (1,2) and (1,3) pairs,
and so on.
By using the results of~\cite{dm,ds}  also these
multiparticle form factors can be evaluated exactly.

\begin{table}[h]
\label{tab4}
\caption{\sl Overlap amplitudes for the spin operator. }
\vskip 0.2cm
\begin{tabular}{|c|}\hline
$ F^{\sigma}_1 =-0.64090211 $ \\
$ F^{\sigma}_2 =\,\,\,\, 0.33867436 $ \\
$ F^{\sigma}_3 =-0.18662854 $ \\
$ F^{\sigma}_4 =\,\,\,\, 0.14277176 $ \\
$ F^{\sigma}_5 =\,\,\,\, 0.06032607 $ \\
$ F^{\sigma}_6 =-0.04338937 $ \\
$ F^{\sigma}_7 =\,\,\,\, 0.01642569 $ \\
$ F^{\sigma}_8 =-0.00303607 $ \\ \hline
\end{tabular}
\end{table}

\begin{table}[h]
\label{tab5}
\caption{\sl Overlap amplitudes for the energy operator. }
\vskip 0.2cm
\begin{tabular}{|c|}\hline
$ F^{\var}_1 =-3.70658437 $ \\
$ F^{\var}_2 =\,\,\,\, 3.42228876 $ \\
$ F^{\var}_3 =-2.38433446 $ \\
$ F^{\var}_4 =\,\,\,\, 2.26840624 $ \\
$ F^{\var}_5 =\,\,\,\, 1.21338371 $ \\
$ F^{\var}_6 =-0.96176431 $ \\
$ F^{\var}_7 =\,\,\,\, 0.45230320 $ \\
$ F^{\var}_8 =-0.10584899 $ \\ \hline
\end{tabular}
\end{table}

As discussed above, in eq.s~(\ref{ffss}), (\ref{ffes}) and (\ref{ffee})
we expect systematic errors of the order 
${\cal O}\left(e^{-2 m_1 r}\right)$. For distances larger than the correlation
length these deviations are very small but become increasingly relevant as the
correlation length is approached. It would be important to have an estimate of
the magnitude of these corrections. From this point of view the present case 
is a perfect laboratory since we know that all other possible
sources of systematic errors are under control.
We report an example of the results obtained using
eq.s~(\ref{ffss}), (\ref{ffes}) and (\ref{ffee})
(for the value $h_l=4.4069\times10^{-4}$)
in the last
column of tabs. 7, 8 and 9.
In order to gain further insight on the convergence of the approximation 
we have also evaluated the contribution of the first eight terms of the
spectral series (i.e. with $m_r=3m_1$). We report in fig.10 the result of this
analysis, together with those with $m_r=2m_1$ and $m_r=m_1$ for comparison, in
the particular case of the $\sigma\sigma$ correlator.

\subsection{IRS approach}

By using the results of sect.~3 and 4 and the exact knowledge of
the constants $R_\sigma$
and $R_\epsilon$ one can easily write the constants
$B_{\Phi\Phi}^i$ in lattice units. They are reported in tab.6\footnote{In 
doing this
conversion one must also take into account the $h$ factor contained in $t$. To
stress this fact we introduce in analogy to $h_l$ a new scaling variable
$t_l\equiv|h_l|~|r|^{15/8}$. In the coefficients reported in tab.6 the
conversion from $t$ to $t_l$ has already been taken into account, hence they 
refer to the IRS expansion in powers of $t_l$ of the correlators}.

We report an example of our results 
(obtained by plugging the values of tab.6 in 
eqs.(\ref{ss}), (\ref{ee}) and (\ref{se}) in the particular case
 $h_l=4.4069\times10^{-4}$), 
in the second
column of tabs. 7, 8 and 9~.

\begin{table}[h]
\label{tab6}
\caption{\sl Coefficients of the IRS expansion in lattice units. }
\vskip 0.2cm
\begin{tabular}{|l|}
\hline
$ B^1_{\sigma\sigma} =0.7034... $ \\
$ B^2_{\sigma\sigma} =0.6414... $ \\
$ B^3_{\sigma\sigma} =-0.3007... $ \\
\hline
$ B^1_{\epsilon\epsilon} =0.1013... $ \\
$ B^2_{\epsilon\epsilon} =0 $ \\
$ B^3_{\epsilon\epsilon} =3.4776... $ \\
\hline
$ B^1_{\sigma\epsilon} =0.1685... $ \\
$ B^2_{\sigma\epsilon} =0.7380... $ \\
$ B^3_{\sigma\epsilon} =-0.3712... $ \\
\hline
\end{tabular}
\end{table}

\subsection{Comparison with MC results}

\subsubsection{Lattice artifacts}
It is interesting to see that  lattice artifacts are confined to a remarkably
small region of few lattice spacings, which shows a negligible dependence on 
the magnetic field or on the type of correlator. 
Since the lattice artifacts decrease so quickly it is rather
easy to find where does  the region of applicability of the IRS results
starts, by simply looking at the distance 
$L_{min}$ where for the first time the IRS prediction becomes
 compatible (within the errors) with the MC data or, if this never happens,
looking at the location of the  
minimum difference between IRS predictions and the MC 
simulations~\footnote{We may expect that higher orders in the $t$ expansion
improve the large distance behaviour of the IRS results, but they give a
negligible contribution around $L_{min}$ where the discrepancy between
Montecarlo data and IRS predictions is completely dominated by lattice 
artifacts.}. It turns out that for all correlators and $h_l$ values $L_{min}$
ranges between 7 and 9 lattice spacings.
This tells us that, at least in the case of the Ising model perturbed by a
magnetic field, the IRS method has a  large window of applicability,
which becomes larger and larger as the critical point is approached.
This is well exemplified by fig.6 and 7 where the difference between MC data 
and IRS predictions is plotted, for the $\sigma\sigma$ correlator
 in the short range region,
 first in units of the lattice spacing and then in units of the correlation
 length.
It would be interesting to test if this behaviour
 also holds for other models or for
different realizations of this one.

\subsubsection{$h_l$ dependence}
In the range $L_{min} < r < \xi$ the agreement between IRS predictions and MC
results is always very good. In particular, it seems that the 
 method reaches its better results in the case of
$\epsilon\sigma$ correlator. As expected,
 the IRS approximation becomes better
and better as we approach the critical point (see fig.s 3, 4 and 5).
 First  because the range of
validity becomes larger and second because the systematic deviations
 due to the terms neglected in the expansion, which are proportional to 
higher powers of $h_l$ become less important.
 In particular for  the $\epsilon\sigma$
correlator at $h_l=1.1017\times10^{-4}$ there is a wide range (more than 40
lattice spacings) in which the IRS prediction coincides with the MC results
within the errors (see fig.8).

\subsubsection{IRS versus FF.}
 Looking at  tables 7, 8, 9 and at the fig.s 1-5 we see that,
 as expected, the FF approach performs better than the IRS one for distances
 larger than the correlation length and that the opposite is true for distances
 smaller than $\xi$. It is interesting to see that for distances of the order of
 the correlation length  the IRS and FF methods give
 comparable performances. Some interesting informations on the
 systematic errors involved in the two approximations can be extracted from the
 data.
  \begin{description}
   \item{1]} While the systematic errors in the IRS approach have a polynomial
    behaviour in the distance $r$ (which is contained in $t_l$), those of the FF
    approach have an exponential behaviour. This is clearly visible in fig.2
    where the deviations are plotted as a function of $r$ for the
correlator $G_{\sigma\sigma}$ at $h_l=4.4069\times 10^{-4}$ and in fig.3-5
where they are plotted for all the correlators and all the values of $h_l$ as
functions of $r/\xi$.
 This makes the IRS
    method still reasonably reliable even for distances twice the correlation
    length.
   \item{2]} We may obtain a rough estimate of the magnitude of the systematic
   errors involved in the IRS approach with the following argument.
Since $t_l\equiv|h_l|~|r|^{15/8}$, looking at eq.(\ref{xi}) we see that for
distances of the order of the correlation length we have $t_l\sim 0.06$. 
Depending on the correlator chosen, we would expect deviations of the order 
$O(t_l^2)$, $O(t_l^{31/15})$, $O(t_l^{32/15})$. 
Since in general the $B_{\Phi\Phi}^i$ constants are of order unity,
this amounts to an expected deviation for the $F_{\Phi\Phi}$ functions 
of order $\delta F\sim 0.004$. 
This expectation is in good agreement  with the values of $\delta F$ obtained by
comparing the MC and IRS estimates at the distance $r=\xi$.
(see for instance the data reported in 
tab.s 7,8 and 9. In using these data one must
take into account the normalization between $F_{\Phi,\Phi}$ and
$G_{\Phi,\Phi}$ functions).

\item{3]}
A similar analysis can be performed in the case of the FF method. In
this case we know that
the systematic errors 
are of order $O(e^{-m_rr})$. We shall further discuss them in sect.6.3.5 below.
 In the large distance regime $r\geq 1.5~\xi$ the performances of the FF
 approach are very good. For instance, in this region, 
for the lowest value of $h$ that we
 studied:
 $h_l=1.1017\times10^{-4}$  
 the FF predictions for the $\epsilon \epsilon$ correlator
coincide with the MC results within the errors (see fig.5).

\end{description}

\subsubsection{Convergence of the IRS expansion}
It is interesting to study the convergence properties of the IRS method, i.e.
to see if the agreement with the Montecarlo data improves as
higher terms are added in the expansion. 
\begin{itemize}
\item \hskip 0.5cm
$\br \epsilon \sigma \kt$ and $\br \sigma \sigma \kt$.

In these two cases the agreement improves as new terms are added in the
expansion. This is clearly visible in fig.8
where we have plotted the
difference between the MC data for the  correlator $\br \epsilon \sigma \kt$
at $h_l=1.1017\times10^{-4}$ and the IRS results with one (pluses), two
(crosses) and three (diamonds) terms in the expansion. The analogous plot for
the $\br \sigma \sigma \kt$ correlator shows exactly the same behaviour.

\item \hskip 0.5cm
$\br \epsilon \epsilon \kt$.

In this case we find exactly the opposite behaviour. As it is shown in fig.9
the new coefficient that we evaluated does $not$ improve the agreement with
the Montecarlo data which is, by the way, impressively good already with the
simple zero order contribution to the correlator.
 We see two possible reasons for this, rather unexpected, behaviour.
\begin{description}
\item{1]}
As noticed in sect.4, the next term in the perturbative expansion of the 
correlator has an exponent $32/15$, which is very near to the one that we
evaluated. We cannot evaluate this further contribution
 since it would require the knowledge of the expectation value
of the $T\bar T$ operator. In principle this term could well compensate 
the deviation that we observe. 
\item{2]} 
This behaviour could be an indication of the bad convergence properties of the
IRS method. If this is the case it would be very interesting to test (in view
of the good behaviour of the other two correlators) if this is a peculiar 
feature of the  $\br \epsilon \epsilon \kt$ correlator or a feature of the
expansion itself. This issue could be settled in principle by looking to the
higher perturbative terms in the expansions of the two other correlators. We
plan to address this point in a forthcoming paper.

\end{description}
\end{itemize}

\subsubsection{Convergence of the FF expansion}
In order to study the convergence of the FF approximation we have compared
the Montecarlo data in the case of the $\br \sigma \sigma \kt$ correlator
at $h_l=1.1017\times10^{-4}$ with the result of the FF approximation truncated
at $m_r=m_1$, $m_r=2m_1$, $m_r=3m_1$ respectively.
 This corresponds to take into account one, three and eight states
 respectively in the spectral sum. The results of the comparison are reported
 in fig.10~. Looking at this figure one may see that as higher orders are added
 the approximation smoothly converges to the MC data. By comparing the
 three approximations one may get a perception of the convergence rate of the
 method.

\section{Concluding remarks}

In this paper we have compared the predictions of the IRS and FF approximations
for the $\sigma\sigma$, $\epsilon\sigma$ and $\epsilon\epsilon$ correlators
with the results of a set of high precision MC simulations of
the 2d Ising model perturbed by a magnetic field. To this end we have extended 
the IRS approach to  second order derivatives of the structure constants.
Our main results are:
\begin{itemize}
\item
Lattice artifacts are confined in a small region of few lattice spacings.
\item
There is a wide region ranging from $\sim 7-9$ lattice spacings to the
correlation length in which the MC data are in good agreement with the IRS
results. 
\item
The agreement improves as the critical point is approached.
\item
For distances smaller than $\xi$ the IRS gives a better approximation than the
FF method, while the opposite is true for distances larger than $\xi$.

\end{itemize}

The IRS method can be extended in principle to any order in the derivatives of
the Wilson coefficients, by using the integration method of~\cite{mathu} and
the technique of the Mellin transform. However in the case that we studied, 
i.e. the $\br \epsilon \epsilon \kt$ correlator,
the IRS method turns out to show rather bad convergence properties. It remains
an open problem to understand if this is a limit of the method itself
 or if it is a peculiar feature of the correlator that we have chosen.

It would be very interesting to extend this analysis to other models in this
same universality class. In particular one could study the model recently
introduced in~\cite{bnw,
gn96} for which an exact bethe ansatz solution, out of the
critical point exists. Another interesting application of the method would be
the study of the correlators in the case of the most general perturbation of
 the Ising critical point (i.e. a mixed situation with both magnetic and thermal
 perturbations). In this case the exact integrability is lost but the
 IRS method is still valid and could give important informations on the
 behaviour of the correlators. In particular it would allow us to compare our
 approximation with the interesting results, directly obtained on the lattice,
 in~\cite{is2}.

\section*{Appendix}

The evaluation of the coefficient 
$\wh{\dep_h^2 C^{\mathbf 1}_{\epsilon \epsilon}}$ involves the calculation 
of the integral 
\eq
Z=\int \txrm{d}^2 w  
\vert w \vert^{e} \vert 1-w \vert^{f} {w^*}^r(1-w)^s
\int \txrm{d}^2 z  \vert z \vert^{\alpha} \vert 1-z \vert^{\beta}  
z^n{(1-z)^*}^m
\vert z-w \vert^{\gamma}
\label{genf}
\en
where $n,m,r,s \in \mathbf N$ and $\alpha, \beta , \gamma,e,f \in
\mathbf R$. \\
It is useful to introduce the following theorem 
(see \cite{mathu}, \cite{gm2}). Let us consider an integral of the form 
\eq
I= \int \txrm{d}^2 w  \sum_{\alpha,\beta=1}^N f_\alpha (w) Q_{\alpha \beta}
{\bar f}_\beta(w^*)
\label{int1}
\en
where $\{ f_\alpha(w) \}_{\alpha=1,N}^N$, 
$\{ {\bar f}_\beta(w^*) \}_{\beta=1,N}^N$ are two sets of independent  
functions and $Q_{\alpha \beta}$ is a constant matrix. 
Let us assume  that 
${\bar f}_\beta (w^*)$ e $(f_\beta (w))^*$ have the same monodromies, in 
particular the two sets of functions
$f_\alpha(w)$ and $g \equiv ({\bar f}_\beta(w^*) )^*$ must have the same 
branch points $\{ w_k \}^{m+1}_{k=0}$ such that
\eq
0=\vert w_0 \vert <  \vert w_1 \vert < \cdots < \vert w_m \vert  <
\vert w_{m+1} \vert =\infty 
\en
and they have to be analytic elsewhere. \\
If we assume now that the matrix $Q$ is invariant under the monodromy
group action
\eq
Q \ = \ M^t_k Q M^*_k, \ \ \ \ \ \ \ \ \forall k
\en
where $M_k$ are the monodromy matrices of $f$ and $g$ related to the branch
points $w_k$,
it follows that we are able to express $I$ in terms of
one-dimensional integrals (see \cite{mathu}, \cite{gm2} for more details)
\eq
I= \frac{i}{2} \sum_{k=1}^m {\mathcal I}^{(k)}_\alpha
\left\lb \left( (1-M_{k+1})^{-1} - (1-M_{k})^{-1} 
\right)^t Q \right\rb_{\alpha \beta}
{\bar{\mathcal I}}^{(k)}_\beta 
\label{a.7}
\en
where $^t$ is the transposition and 
\eqa
{\mathcal I}^{(k)} & \equiv & \int_{C_k} \txrm{d} w f(w)  
\nonu\\
{\bar{\mathcal I}}^{(k)} & \equiv & \int_{{\bar C}_k} \txrm{d} w^* 
{\bar f}(w^*)  
\ena
where $C_k$ (${\bar C}_{k}$) are counter-clockwise (clockwise) 
circumferences enclosing all the branch points of  modulus lower than $w_{k}$,
starting at $w_{k_+}$ (infinitesimally over the cut at $w_k$) and ending
at $w_{k_-}$ (infinitesimally under the cut at $w_k$).
\vskip0.4cm
Now we are able to evaluate both the $z$-plane and $w$-plane integrations
of (\ref{genf}) using the previous lemma. \\
First, to perform the $z$-plane integration, we pose
\eq
I_z(w,w^*) =
\int \txrm{d}^2 z  \vert z \vert^{\alpha} \vert 1-z \vert^{\beta}  
z^n{(1-z)^*}^m
\vert z-w \vert^{\gamma}
\en
so the $z$-plane integration involves the following branch points
\eq
z_0 = 0, \ \ \ z_1 = w, \ \ \ z_2 = 1, \ \ \ z_\infty = \infty.
\en
Thus, the application of (\ref{a.7}) gives
\eq
I_z (w,w^*)=\frac{i}{2} \left\lb 
{\mathcal I}^{(1)}_1 T^{(1)}_{12} {\bar{\mathcal I}}^{(1)}_2 
\right\rb +
\frac{i}{2} \left\lb 
{\mathcal I}^{(2)}_1 T^{(2)}_{12} {\bar{\mathcal I}}^{(2)}_2 
\right\rb 
\label{idiz}
\en
where
\eq
T^{(1)}_{12} \ = \ 
\frac{e^{i \pi \alpha}(e^{i \pi \gamma}-1)}
{(e^{i \pi (\alpha+\gamma)}-1)(e^{i \pi \alpha}-1)}
\en
and
\eq
T^{(2)}_{12} \ = \ 
\frac{e^{i \pi (\alpha+\gamma)}(e^{i \pi \beta}-1)}
{(e^{i \pi (\alpha+\beta+\gamma)}-1)(e^{i \pi (\alpha+\gamma)}-1)} 
\en
are the only non vanishing entries of the matrices
\eqa
T^{(1)} & = & 
\left( (1-M_{2})^{-1} - (1-M_{1})^{-1} 
\right)^t Q
\nonu\\
T^{(2)} & = & 
\left( (1-M_{\infty})^{-1} - (1-M_{2})^{-1} 
\right)^t Q.
\ena
This imply that we have to take in account only the integrals
\eqa
{\mathcal I}^{(1)}_1 & = & (e^{-i \pi \alpha}-1)  w^{1+\alpha/2+\gamma/2+n}
\frac{\Gamma(\alpha/2+n+1)\Gamma(\gamma/2+1)}{\Gamma(\alpha/2+\gamma/2+2+n)}    
\cdot
\nonu\\
& \cdot  & F(-\beta/2,\alpha/2+n+1;\alpha/2+\gamma/2+2+n;w);
\nonu\\
\bar {\mathcal I}^{(1)}_2 & = & (e^{i \pi \alpha}-1)  
{w^*}^{1+\alpha/2+\gamma/2}
\frac{\Gamma(\alpha/2+1)\Gamma(\gamma/2+1)}{\Gamma(\alpha/2+\gamma/2+2)}    
\cdot
\nonu\\
& \cdot  & F(-m-\beta/2,\alpha/2+1;\alpha/2+\gamma/2+2;w^*)
\ena
and
\eqa
{\mathcal I}^{(2)}_1  
& = &  (e^{-i \pi (\alpha+ \beta+ \gamma)}-1)
(-)^m \frac{\Gamma(-\alpha/2-\beta/2-\gamma/2-n-1)
\Gamma(\beta/2+1)}{\Gamma(-\alpha/2-\gamma/2-n)}    
\cdot
\nonu\\
& \cdot &  F(-\gamma/2,-\alpha/2-\beta/2-\gamma/2-n-1;
-\alpha/2-\gamma/2-n;w);
\nonu\\
{\bar{\mathcal I}}^{(2)}_2 
& = & 
(e^{i \pi (\alpha+ \beta+ \gamma)}-1)
\frac{\Gamma(-\alpha/2-\beta/2-\gamma/2-m-1) \Gamma(\beta/2+1+m)}
{\Gamma(-\alpha/2-\gamma/2)} 
\cdot
\nonu\\
& \cdot &  F(-\gamma/2,-\alpha/2-\beta/2-\gamma/2-m-1;
-\alpha/2-\gamma/2;w^*).
\ena
Finally, putting all these relations in (\ref{idiz}), we can recover the 
wanted result for $I_z (w,w^*)$.
\vskip0.4cm
The $w$-plane integration is very similar to the previous one.
Now we have to evaluate 
\eq
Z=\int \txrm{d}^2 w  
\vert w \vert^{e} \vert 1-w \vert^{f} {w^*}^r(1-w)^s
\ I_z(w,w^*)
\label{intw}
\en
which involves $w_0=0$, $w_1=1$, $w_\infty=\infty$ as branch points.
Hence the solution is given by (\ref{a.7}), i.e.
\eq
Z=\frac{i}{2} \left\lb 
{\mathcal I}^{(1)} T^{(1)} {\bar{\mathcal I}}^{(1)} 
\right\rb
\en
where 
\eq
T^{(1)} =  
\left( (1-M_{\infty})^{-1} - (1-M_{0})^{-1} 
\right)^t Q.
\en
The contribution coming from ${\mathcal I}^{(1)}$ is
\eqa
{\mathcal I}^{(1)}_1 & = &  
(e^{i \pi e}-1) \int_0^1 \txrm{d} z f_1 = (e^{i \pi e}-1) \ J_1     
\nonu\\
{{\mathcal I}}^{(1)}_2 &=&
(e^{-i \pi (e+\alpha+\gamma)}-1) \int_0^1 \txrm{d} z f_2 = 
(e^{-i \pi (e+\alpha+\gamma)}-1) \ J_2     
\nonu\\  
{\bar{\mathcal I}}^{(1)}_1 &=& 
(e^{-i \pi e}-1) \int_0^1 \txrm{d} z^* {\bar f}_1 = (e^{-i \pi e}-1) \ 
{\bar J}_1     
\nonu\\  
{\bar{\mathcal I}}^{(1)}_2 &=& 
(e^{i \pi (e+\alpha+\gamma)}-1) \int_0^1 \txrm{d} z^* {\bar f}_2 = 
(e^{i \pi (e+\alpha+\gamma)}-1)\ {\bar J}_2     
\ena
that, in terms of generalized hypergeometric functions, becomes
\eqa
 J_1 &=& B(-\alpha/2-\beta/2-\gamma/2-n-1,\beta/2+1)
B(1+e/2,1+f/2+s)
\nonu\\
&&_3F_2(\frac{-\alpha-\beta-\gamma}{2}-n-1,-\frac{\gamma}{2},1+\frac{e}{2};
\frac{-\alpha-\gamma}{2}-n,2+\frac{e+f}{2}+s);1)
\nonu\\
J_2 &=& B(1+\gamma/2,1+\alpha/2+n)
B(2+e/2+\alpha/2+\gamma/2+n,1+f/2+s)
\nonu\\
&&_3F_2(-\frac{\beta}{2},2+\frac{\alpha+\gamma}{2}+n+e/2,1+\frac{\alpha}{2}+n;
2+\frac{\alpha+\gamma}{2}+n,3+
\nonu\\
&&+ \frac{\alpha+\gamma+e+f}{2}+n+s);1)
\nonu\\
 {\bar J}_1 &=& B(-\alpha/2-\beta/2-\gamma/2-m-1,\beta/2+1+m)
B(1+e/2+k,1+f/2)
\nonu\\
&&_3F_2(\frac{-\alpha-\beta-\gamma}{2}-m-1,\frac{-\gamma}{2},1+e/2+k;
\frac{-\alpha-\gamma}{2},2+\frac{e+f}{2}+k);1)
\nonu\\
 {\bar J}_2 &=& B(1+\gamma/2,1+\alpha/2)
B(2+e/2+\alpha/2+\gamma/2+k,1+f/2)
\nonu\\
&&_3F_2(\frac{-\beta}{2}-m,2+\frac{\alpha+\gamma+e}{2}+k,1+\frac{\alpha}{2};
2+\frac{\alpha+\gamma}{2},3+
\nonu\\
&& + \frac{\alpha+\gamma+e+f}{2}+k);1).
\nonu\\
\ena
Thus the solution has the form 
\eq
Z \ = \ t_{11} J_1 {\bar J}_1 + t_{12} J_1 {\bar J}_2 +
t_{21} J_2 {\bar J}_1 + t_{22} J_2 {\bar J}_2
\en
where the matrix elements $t_{ij}$ are the following
\eqa
t_{11} & = & \Delta^{-1} S(e/2)S(\beta /2)S((\alpha +\beta + \gamma )/2)  
\cdot \nonu \\
& \cdot & \bigg( S(\alpha /2)S(\beta /2)S(f/2)  S((\alpha +\beta +e+f+
2\gamma )/2) +  \nonu \\
& + & S(\gamma /2)S((\alpha + \beta +\gamma )/2) S((\alpha +e+f+\gamma )/2)
S((f+ \gamma +\beta )/2) \bigg)   
\nonu
\ena
\eqa
t_{22} & = & \Delta^{-1} S(\alpha /2)S(\gamma /2)S((\alpha +e+ \gamma )/2)  
\cdot \nonu \\
&\cdot& \bigg( S(\alpha /2)S(\beta /2)S((e+f)/2)S(1/2(\beta +f+\gamma ))+ 
\nonu \\
&+& S(\gamma /2)S((\alpha +\beta +\gamma )/2) S(f/2)S((e+f+\gamma + 
\beta)/2)) \bigg)    
\nonu
\ena
\eqa
t_{12} & = & t_{12} =\Delta^{-1} S(\alpha /2 )S(\beta /2 )S(e/2)
S(\gamma /2) 
\cdot \nonu \\
& \cdot & S((\alpha +e+\gamma )/2)
S((\alpha +\beta +\gamma)/2)S((\beta +\gamma )/2)  
\nonu
\ena 
with
\eqa
\Delta & = & S((\alpha +\gamma )/2)  
\nonu \\
& \cdot & \bigg( (S(\alpha /2)S(\beta /2)S((e+f)/2)S((\alpha +\beta +
e+f+2 \gamma )/2) + 
\nonu \\
& + & S(\gamma /2)S((\alpha + \beta +\gamma )/2)
S((\alpha +e+f+\gamma )/2)S((e+f+\gamma +\beta )/2) \bigg)
\nonu
\ena
and $S(x)=\sin(\pi x)$. \\
For all details on the calculation we refer to \cite{pt}.

\vskip 1cm
{\bf  Acknowledgements}
We thank  F.Gliozzi, R.Guida, M. Hasenbusch and R.Tateo for helpful 
discussions and for a critical reading of the paper. 
This work was partially supported by the 
European Commission TMR programme ERBFMRX-CT96-0045.

\newpage

\begin{table}[h]
\label{tab7}
\caption{\sl Comparison of  the Montecarlo
estimates (second column), the IRS results (third column) 
and the form factor (FF) results (fourth column) for the
correlator $G_{\sigma\sigma}$ at $h_l=4.4069\times 10^{-4}$.
In the first column is reported the distance in lattice units.}
\vskip 0.2cm
\begin{tabular}{|c|c|c|c|}
\hline
$r$ & MC  & IRS & FF     \\
\hline
    1 & 0.71643( 2) & 0.71371 & 0.59385 \\ 
    2 & 0.61138( 3) & 0.60871 & 0.54528 \\ 
    3 & 0.55862( 4) & 0.55764 & 0.51757 \\ 
    4 & 0.52628( 5) & 0.52590 & 0.49852 \\ 
    5 & 0.50401( 6) & 0.50385 & 0.48425 \\ 
    6 & 0.48757( 6) & 0.48750 & 0.47303 \\ 
    7 & 0.47487( 7) & 0.47483 & 0.46393 \\ 
    8 & 0.46475( 7) & 0.46472 & 0.45638 \\ 
    9 & 0.45651( 7) & 0.45647 & 0.45002 \\ 
   10 & 0.44968( 8) & 0.44961 & 0.44459 \\ 
   11 & 0.44393( 8) & 0.44383 & 0.43991 \\ 
   12 & 0.43904( 8) & 0.43889 & 0.43584 \\ 
   13 & 0.43484( 9) & 0.43463 & 0.43229 \\ 
   14 & 0.43120( 9) & 0.43093 & 0.42916 \\ 
   15 & 0.42803( 9) & 0.42768 & 0.42639 \\ 
   16 & 0.42526(10) & 0.42481 & 0.42394 \\ 
   17 & 0.42281(10) & 0.42226 & 0.42175 \\ 
   18 & 0.42065(10) & 0.41998 & 0.41980 \\ 
   19 & 0.41874(10) & 0.41792 & 0.41805 \\ 
   20 & 0.41702(10) & 0.41606 & 0.41648 \\ 
   21 & 0.41549(11) & 0.41436 & 0.41506 \\ 
   22 & 0.41412(11) & 0.41280 & 0.41378 \\ 
   23 & 0.41290(11) & 0.41137 & 0.41263 \\ 
   24 & 0.41179(11) & 0.41003 & 0.41158 \\ 
   25 & 0.41079(11) & 0.40879 & 0.41064 \\ 
   26 & 0.40988(12) & 0.40762 & 0.40978 \\ 
   27 & 0.40907(12) & 0.40652 & 0.40899 \\ 
   28 & 0.40832(12) & 0.40547 & 0.40828 \\ 
   29 & 0.40765(12) & 0.40447 & 0.40763 \\ 
   30 & 0.40703(12) & 0.40351 & 0.40704 \\ 
\hline
\end{tabular}
\end{table}
\newpage

\begin{table}[h]
\label{tab8}
\caption{\sl Comparison of  the Montecarlo
estimates (second column), the IRS results (third column) 
and the form factor (FF) results (fourth column) for the
correlator $G_{\epsilon\epsilon}$ at $h_l=4.4069\times 10^{-4}$.
In the first column is reported the distance in lattice units.}
\vskip 0.2cm
\begin{tabular}{|c|c|c|c|}
\hline
$r$ & MC  & IRS & FF     \\
\hline
    1 & 0.104067( 2) & 0.101321 & 0.002330 \\ 
    2 & 0.029348( 2) & 0.025332 & 0.001735 \\ 
    3 & 0.012327( 3) & 0.011262 & 0.001399 \\ 
    4 & 0.006674( 2) & 0.006340 & 0.001169 \\ 
    5 & 0.004190( 2) & 0.004064 & 0.000999 \\ 
    6 & 0.002879( 2) & 0.002829 & 0.000867 \\ 
    7 & 0.002103( 2) & 0.002087 & 0.000761 \\ 
    8 & 0.001606( 2) & 0.001608 & 0.000674 \\ 
    9 & 0.001268( 2) & 0.001281 & 0.000601 \\ 
   10 & 0.001029( 2) & 0.001049 & 0.000540 \\ 
   11 & 0.000854( 2) & 0.000880 & 0.000488 \\ 
   12 & 0.000720( 2) & 0.000753 & 0.000443 \\ 
   13 & 0.000615( 2) & 0.000657 & 0.000404 \\ 
   14 & 0.000534( 2) & 0.000582 & 0.000371 \\ 
   15 & 0.000469( 2) & 0.000524 & 0.000341 \\ 
   16 & 0.000415( 2) & 0.000478 & 0.000315 \\ 
   17 & 0.000370( 2) & 0.000442 & 0.000292 \\ 
   18 & 0.000334( 2) & 0.000414 & 0.000272 \\ 
   19 & 0.000305( 2) & 0.000392 & 0.000254 \\ 
   20 & 0.000280( 2) & 0.000375 & 0.000238 \\ 
   21 & 0.000258( 2) & 0.000362 & 0.000224 \\ 
   22 & 0.000239( 2) & 0.000353 & 0.000212 \\ 
   23 & 0.000221( 2) & 0.000347 & 0.000200 \\ 
   24 & 0.000205( 2) & 0.000344 & 0.000190 \\ 
   25 & 0.000194( 2) & 0.000342 & 0.000181 \\ 
   26 & 0.000184( 2) & 0.000343 & 0.000173 \\ 
   27 & 0.000174( 2) & 0.000345 & 0.000165 \\ 
   28 & 0.000166( 2) & 0.000349 & 0.000159 \\ 
   29 & 0.000158( 2) & 0.000354 & 0.000153 \\ 
   30 & 0.000152( 2) & 0.000360 & 0.000147 \\ 
\hline
\end{tabular}
\end{table}
\newpage

\begin{table}[h]
\label{tab9}
\caption{\sl Comparison of  the Montecarlo
estimates (second column), the IRS results (third column) 
and the form factor (FF) results (fourth column) for the
correlator $G_{\sigma\epsilon}$ at $h_l=4.4069\times 10^{-4}$.
In the first column is reported the distance in lattice units.}
\vskip 0.2cm
\begin{tabular}{|c|c|c|c|}
\hline
$r$ & MC  & IRS & FF     \\
\hline
    1 & 0.104077(19) & 0.101011 & 0.025867 \\ 
    2 & 0.053137(12) & 0.050882 & 0.020745 \\ 
    3 & 0.034975(11) & 0.034286 & 0.017834 \\ 
    4 & 0.026275(11) & 0.026062 & 0.015840 \\ 
    5 & 0.021240(11) & 0.021181 & 0.014353 \\ 
    6 & 0.017966(11) & 0.017967 & 0.013190 \\ 
    7 & 0.015676(11) & 0.015704 & 0.012251 \\ 
    8 & 0.013990(10) & 0.014032 & 0.011476 \\ 
    9 & 0.012705(10) & 0.012753 & 0.010826 \\ 
   10 & 0.011699(10) & 0.011748 & 0.010274 \\ 
   11 & 0.010891(10) & 0.010942 & 0.009800 \\ 
   12 & 0.010231(11) & 0.010282 & 0.009391 \\ 
   13 & 0.009685(11) & 0.009736 & 0.009035 \\ 
   14 & 0.009228(11) & 0.009277 & 0.008723 \\ 
   15 & 0.008841(11) & 0.008888 & 0.008448 \\ 
   16 & 0.008510(11) & 0.008555 & 0.008206 \\ 
   17 & 0.008224(11) & 0.008268 & 0.007991 \\ 
   18 & 0.007978(11) & 0.008018 & 0.007800 \\ 
   19 & 0.007765(10) & 0.007800 & 0.007629 \\ 
   20 & 0.007577(10) & 0.007608 & 0.007476 \\ 
   21 & 0.007411(11) & 0.007438 & 0.007340 \\ 
   22 & 0.007265(11) & 0.007287 & 0.007217 \\ 
   23 & 0.007134(11) & 0.007152 & 0.007106 \\ 
   24 & 0.007018(11) & 0.007031 & 0.007006 \\ 
   25 & 0.006914(11) & 0.006922 & 0.006916 \\ 
   26 & 0.006822(11) & 0.006823 & 0.006834 \\ 
   27 & 0.006740(11) & 0.006733 & 0.006760 \\ 
   28 & 0.006667(11) & 0.006650 & 0.006693 \\ 
   29 & 0.006600(12) & 0.006575 & 0.006632 \\ 
   30 & 0.006539(12) & 0.006505 & 0.006577 \\ 
\hline
\end{tabular}
\end{table}

\begin{figure}
\begin {center}
 \null\hskip 1pt
 \epsfxsize 14cm 
 \epsffile{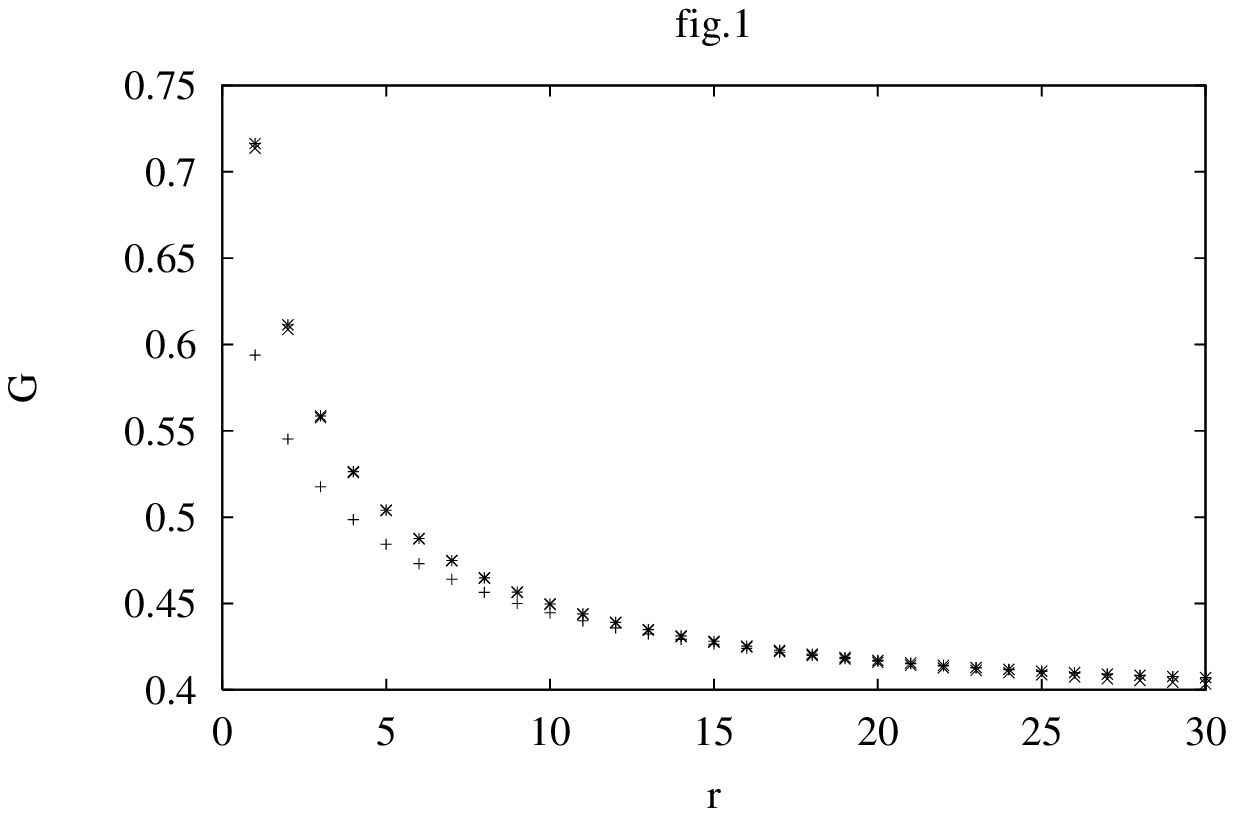}
\end{center}
\mycaptionl{ Comparison of  the Montecarlo
estimates (crosses), the IRS results (asterisks) 
and the form factor results (pluses) for the
correlator $G_{\sigma\sigma}$ at $h_l=4.4069\times 10^{-4}$.}
\end{figure}
\begin{figure}
\begin {center}
 \null\hskip 1pt
 \epsfxsize 14cm 
 \epsffile{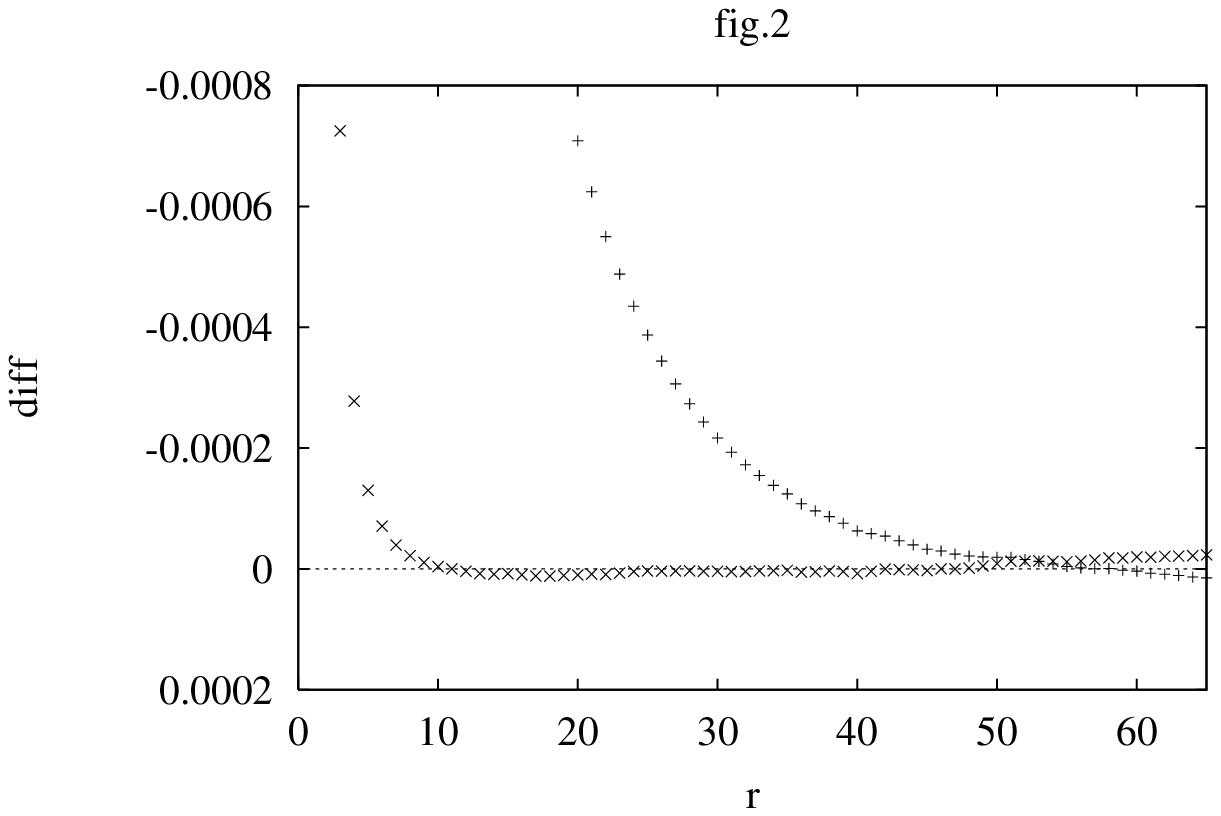}
\end{center}
\mycaptionl{ Differences between  MC
estimates and IRS results (crosses) and between MC
 and form factor results (pluses) for the
correlator $G_{\epsilon\sigma}$ at $h_l=1.1017\times 10^{-4}$.
In this figure, and in all the following ones
the errors in the MC estimates are not reported since they are 
smaller than the symbol size.}
\end{figure}
\begin{figure}
\begin {center}
 \null\hskip 1pt
 \epsfxsize 14cm 
 \epsffile{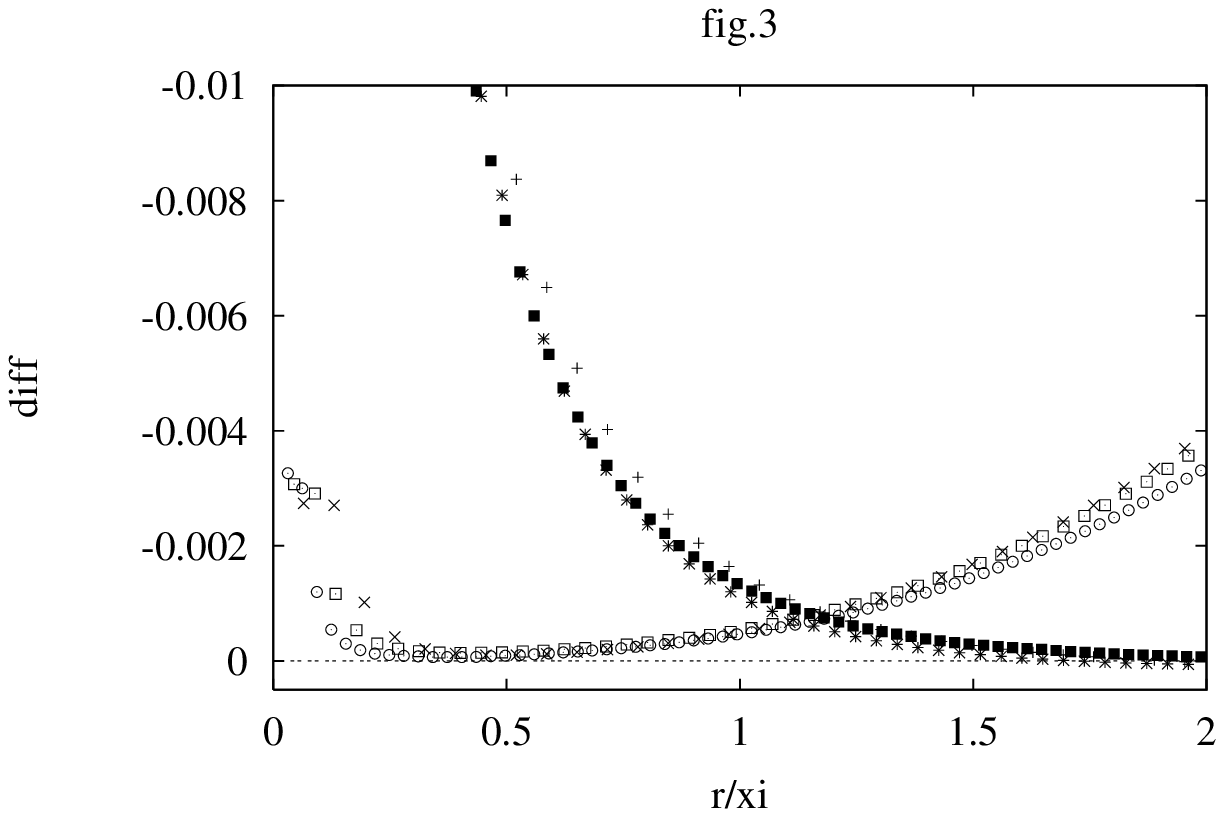}
\end{center}
\mycaptionl{ Differences between  MC
estimates and IRS results (crosses for $h_l=4.4069\times 10^{-4}$,
dotted squares for $2.2034\times10^{-4}$ and circles for
$1.1017\times10^{-4}$) and between MC
 and form factor results (pluses for $h_l=4.4069\times 10^{-4}$,
diamonds for $2.2034\times10^{-4}$ and filled squares for
$1.1017\times10^{-4}$)diamonds) for the
correlator $G_{\sigma\sigma}$. Distances are measured in units of the
correlation length.}
\end{figure}
\begin{figure}
\begin {center}
 \null\hskip 1pt
 \epsfxsize 14cm 
 \epsffile{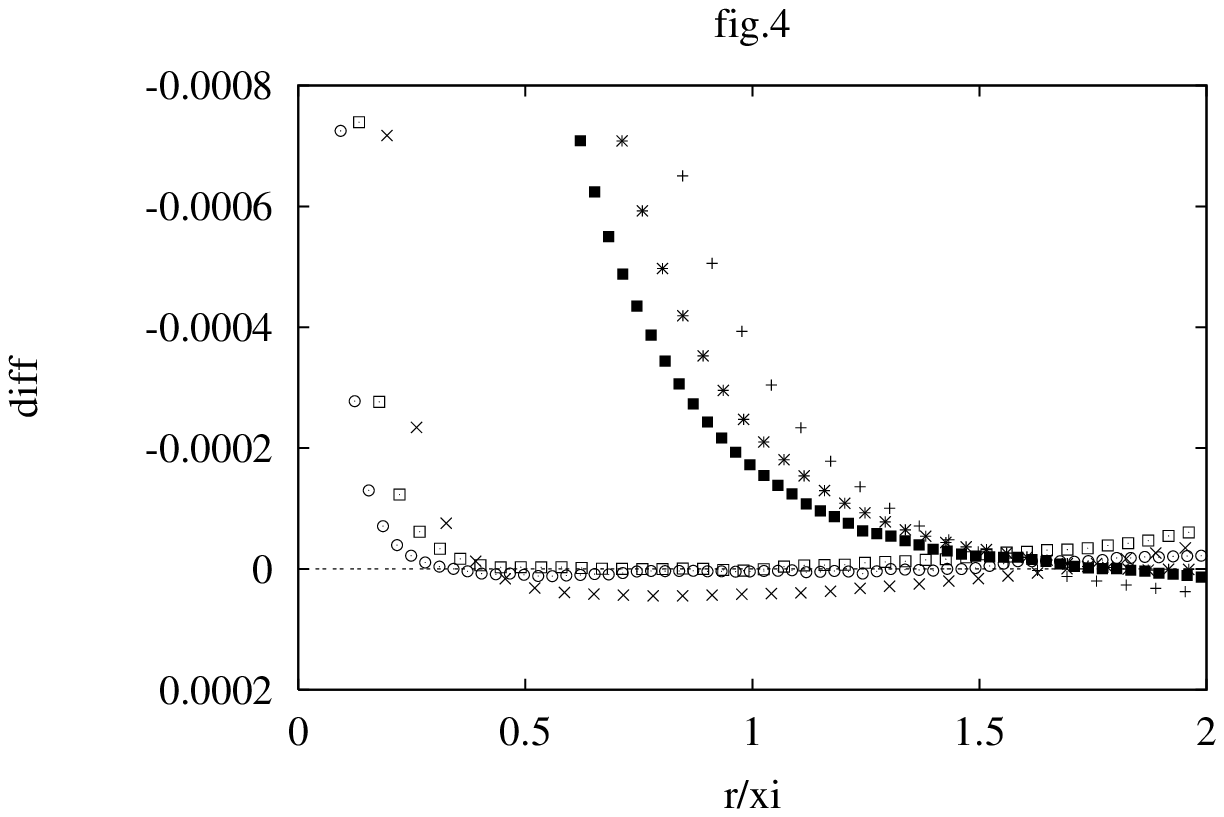}
\end{center}
\mycaptionl{Same as fig.3, but for the $G_{\sigma\epsilon}$ correlator.
Notice the different scale on the $y$ axis}
\end{figure}
\begin{figure}
\begin {center}
 \null\hskip 1pt
 \epsfxsize 14cm 
 \epsffile{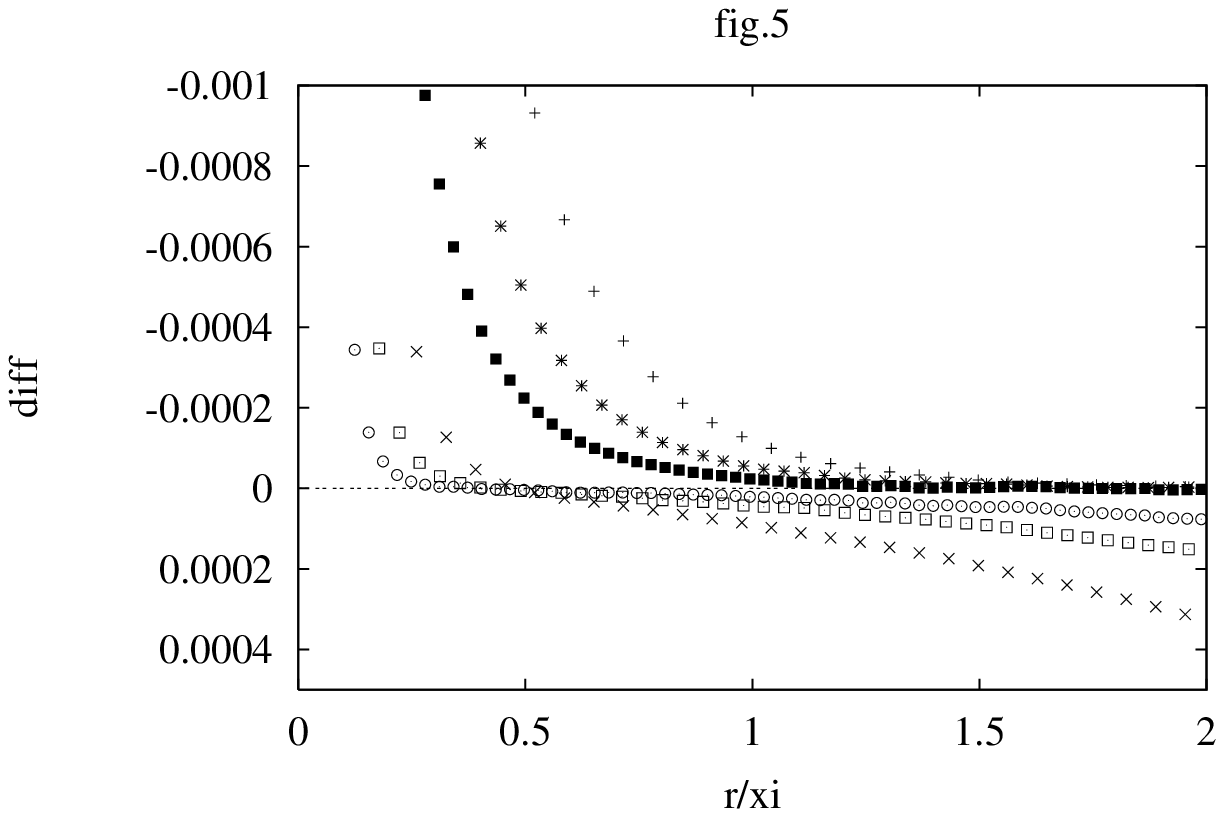}
\end{center}
\mycaptionl{Same as fig.3, but for the $G_{\epsilon\epsilon}$ correlator.}
\end{figure}
\begin{figure}
\begin {center}
 \null\hskip 1pt
 \epsfxsize 14cm 
 \epsffile{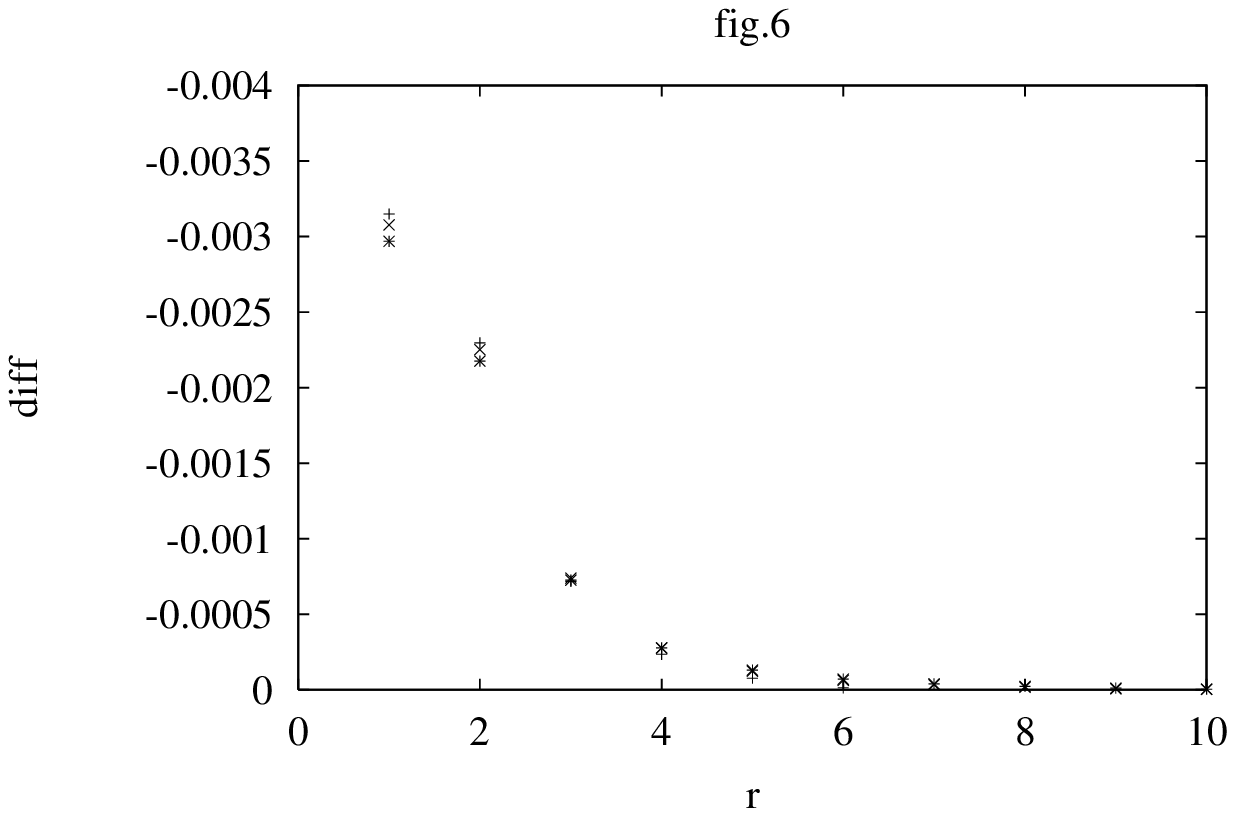}
\end{center}
\mycaptionl{
Differences between  MC
estimates and IRS results for $h_l=4.4069\times 10^{-4}$ (pluses),
$h_l=2.2034\times10^{-4}$ (crosses) and 
$h_l=1.1017\times10^{-4}$ (diamonds) for the $\epsilon\sigma$ correlator, in the
short distance region dominated by the lattice artifacts.}
\end{figure}
\begin{figure}
\begin {center}
 \null\hskip 1pt
 \epsfxsize 14cm 
 \epsffile{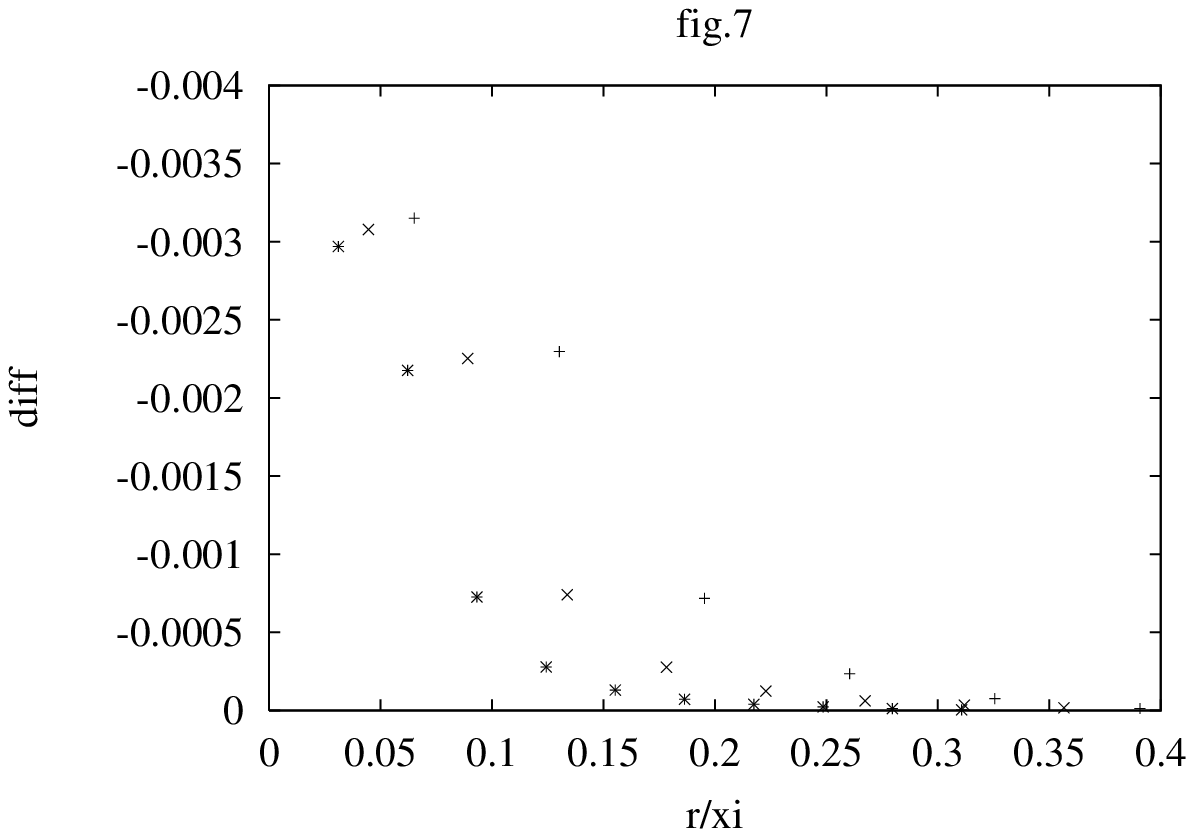}
\end{center}
\mycaptionl{ Same as in fig.6, but with distances measured
 in units of the correlation length.}
\end{figure}
\begin{figure}
\begin {center}
 \null\hskip 1pt
 \epsfxsize 14cm 
 \epsffile{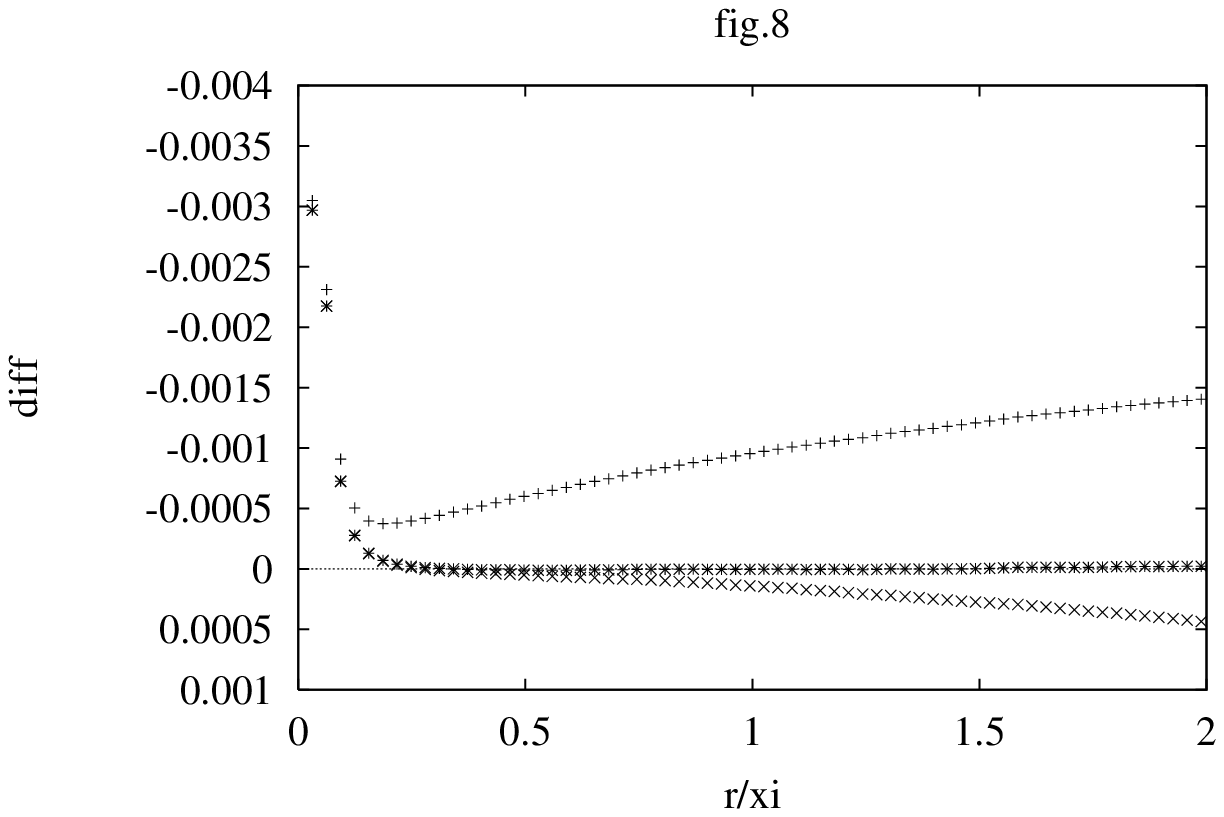}
\end{center}
\mycaptionl{
Difference between the MC data for the $\br \epsilon \sigma \kt$ correlator 
at $h_l=1.1017\times10^{-4}$ and the IRS results with one (pluses), two
(crosses) and three (diamonds) terms in the expansion. }
\end{figure}
\begin{figure}
\begin {center}
 \null\hskip 1pt
 \epsfxsize 14cm 
 \epsffile{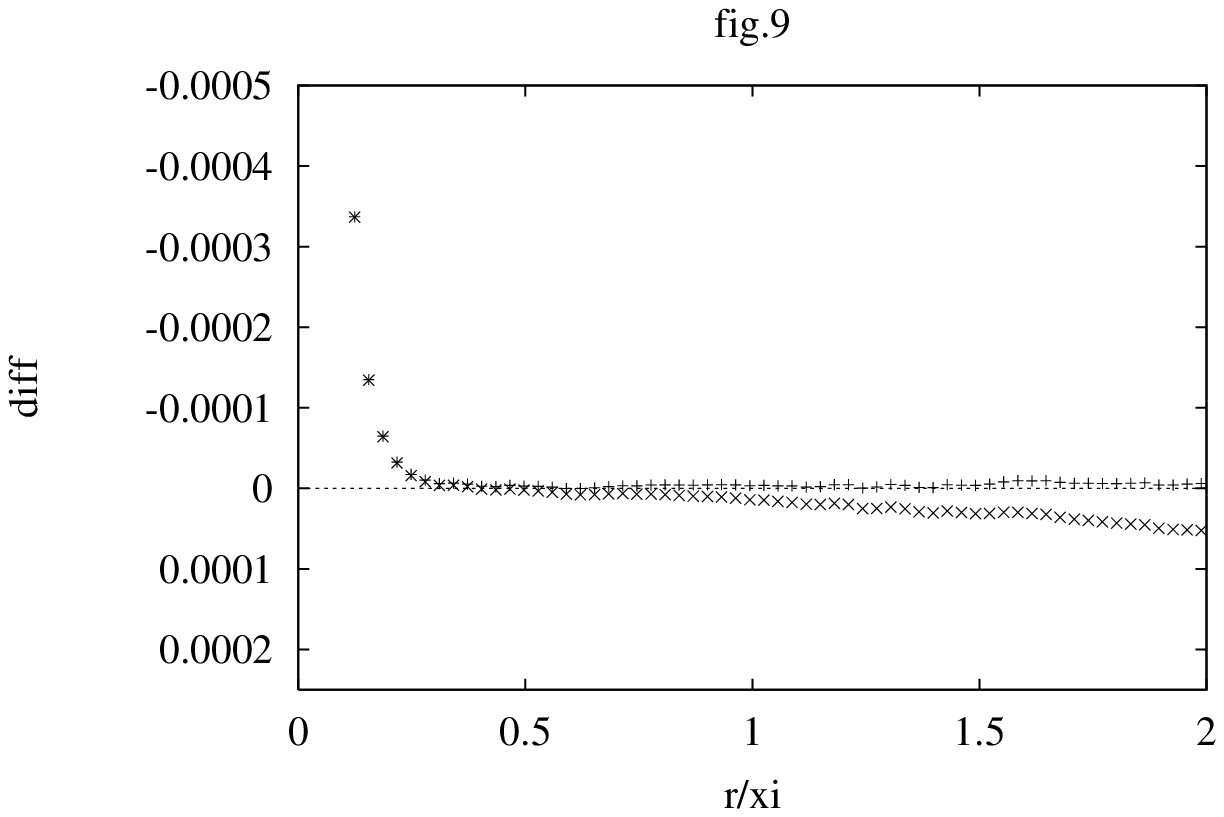}
\end{center}
\mycaptionl{
Difference between the MC data for the $\br \epsilon \epsilon \kt$ correlator 
at $h_l=1.1017\times10^{-4}$ and the IRS results with one (pluses) and two
(crosses) terms in the expansion. }
\end{figure}
\begin{figure}
\begin {center}
 \null\hskip 1pt
 \epsfxsize 14cm 
 \epsffile{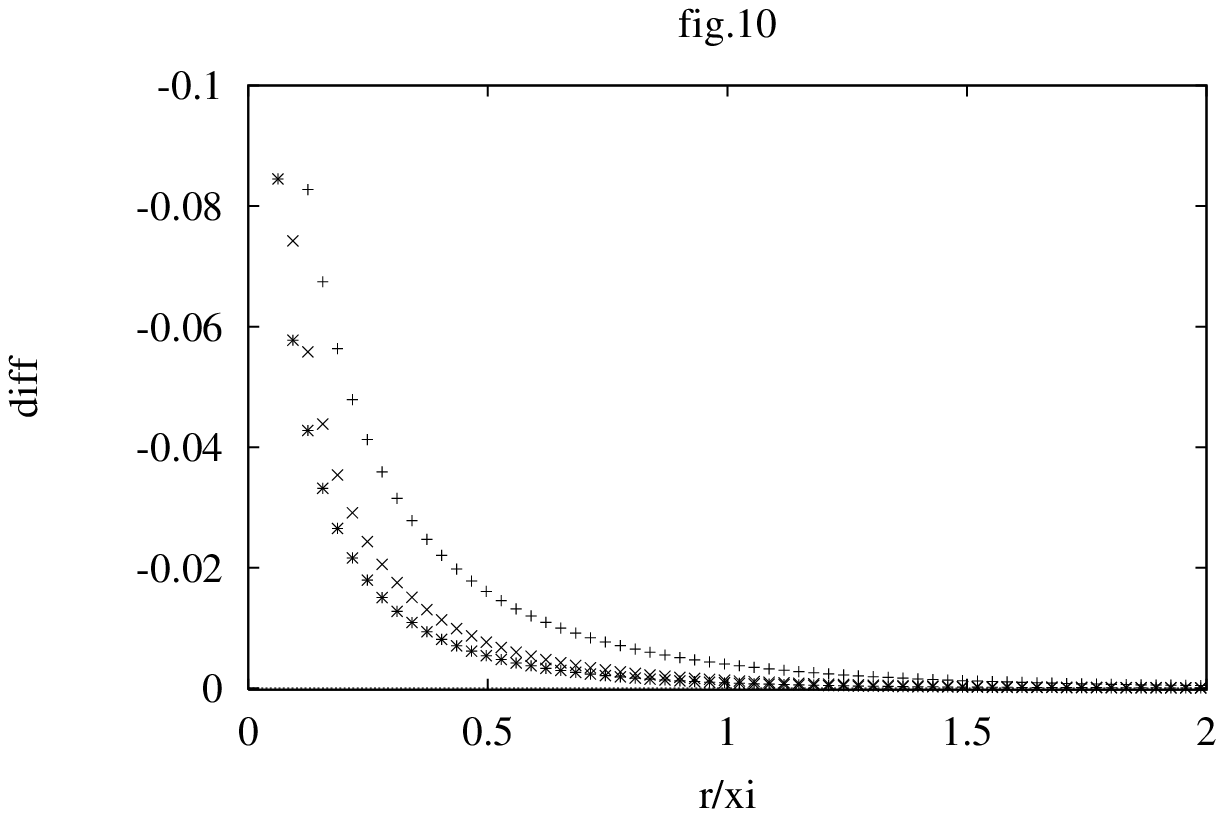}
\end{center}
\mycaptionl{
Difference between the MC data for the $\br \sigma \sigma \kt$ correlator 
at $h_l=1.1017\times10^{-4}$ and the FF results with one (pluses), three
(crosses) and eight (diamonds) terms in the spectral series. }
\end{figure}

\end{document}